\documentclass[active,tightpage]{pasa}%
\title[IFU spectroscopy of Southern Planetary Nebulae]{IFU spectroscopy of Southern Planetary Nebulae V:
Low-ionization structures}
\author[A.Ali \& M.A. Dopita]{A. Ali,$^{1,2}$ \and M.A. Dopita,$^{3}$\thanks{E-mail: afmali@kau.edu.sa; Michael.Dopita@anu.edu.au}\\
\affil{$^1$Astronomy Dept, Faculty of Science, King Abdulaziz University, Jeddah, Saudi Arabia.}%
\affil{$^2$Department of Astronomy, Faculty of Science, Cairo University, 12613, Giza, Egypt.}%
\affil{$^3$Research School of Astronomy and Astrophysics, Australian National University, Cotter Rd., Weston ACT 2611, Australia.}}%
\jid{PASA}
\doi{10.1017/pas.\the\year.xxx}
\jyear{\the\year}
\usepackage{graphicx}
\usepackage{pdflscape}
\usepackage{longtable}
\usepackage{epstopdf}
\usepackage[authoryear]{natbib}
\bibpunct{(}{)}{;}{a}{}{,}
\usepackage{varwidth}
\usepackage{aas_macros}
\usepackage{hyperref}
\hypersetup{colorlinks,citecolor=blue,linkcolor=blue,urlcolor=blue}

\begin{document}%
\begin{abstract}
 In this 5th paper of the series, we examine the spectroscopy and morphology of four southern Galactic planetary nebulae Hen\,2-141,
 NGC\,5307, IC\,2553, and PB\,6  using new integral field spectroscopy data. The morphologies and ionization structures of the sample are given as a set of
 emission-line maps. In addition, the physical conditions, chemical compositions, and kinematical characteristics of these objects are derived.
 The results show that  PB\,6 and Hen\,2-141 are of very high excitation classes  and  IC\,2553 and NGC\,5307 are mid to high excitation objects.
 The elemental abundances reveal that PB\,6 is of Type I, Hen\,2-141 and IC\,2553 are of Type IIa, and NGC\,5307 of Type IIb/III. The observations unveil the
presence of well-defined low-ionization structures or ``knots" in
all objects. The diagnostic diagrams reveal that the excitation
mechanism of these knots is probably by photo-ionization of dense
material by the nebular central stars. The physical analysis of six
of these knots show no significant differences with their
surrounding nebular gas, except their lower electron densities. In
spite of the enhancement of the low-ionization emission lines of
these knots, their chemical abundances are nearly comparable to
their surrounding nebulae, with the exception of perhaps slightly
higher nitrogen abundances in the NGC\,5307 knots. The integrated
spectrum of IC\,2553 reveals that nearly all key lines that have led
researchers to characterize its central star as a weak-emission line
star type are in fact of nebular origin.\end{abstract}
\maketitle%
\section{Introduction}

The study of  planetary nebulae (PNe) and their central stars (CSs) provides valuable constraints on the evolution of both low and
intermediate mass stars from the asymptotic giant branch (AGB) to the white dwarf phase of evolution. Almost all spectroscopic studies of PNe, up to now, have applied long-slit spectroscopic technique in the analysis. Such measurements are confined to a small portion of the entire nebula depending on the slit size and position. By contrast, measurements using the integrated field unit (IFU) technique cover the entire nebula (assuming that its angular size is
smaller than the instrument field of view).

 In this article, we continue our journey to analyze the IFU spectra for southern Galactic PNe using the Wide Field Spectrograph (WiFeS)
 instrument mounted on the 2.3 m ANU telescope at Siding Spring Observatory. Precise nebular analysis and modelling depend on an understanding of
 their integrated spectra and spatial structures. The data cubes generated by the WiFeS device help us not only to study the PNe physical and kinematical
properties, but also to explain the internal flux distributions,
ionization structures, and overall morphologies. More details
regarding the advantages of using the IFU spectroscopy compared to
the long-slit spectroscopy have been presented in earlier papers,
see \citet{Ali16} and \citet{Basurah16}.

The first uses of the IFU spectroscopy in this field was by
\citet{Monreal-Ibero05} and, a little later by \citet{Tsamis07}.
More recently, \citet{Ali16} have used the WiFeS instrument to
extract integrated spectra of the Galactic planetary nebulae: M3-4,
M 3-6, Hen 2-29, and Hen 2-37. These observations allowed extraction
of the spectrum of the central star (CS) in the M 3-6 nebula. This
was revealed to be H-rich and of the spectral type O3 I(f*).
Further, they found most of the recombination lines which used
previously to classify the CS as a weak-emission line star (WELS)
type  arise from the nebula rather than from its CS.
\citet{Basurah16} found another four examples of such
misclassification of PNe CSs as WELS type, and provided detailed
nebular analysis and modeling for the highly excited PNe: NGC 3211,
NGC 5979, My 60, and M 4-2. \citet{Ali15} extracted the integrated
spectrum of the large, evolved, and interacting planetary nebula
(PN) PNG$342.0-01.7$. The full spatial extent of the object required
a mosaic of nine observing WiFeS frames.

The present article sheds light on the morphology and spectroscopy
of another four southern PNe that are associated with the presence
of low-ionization structures (LISs) ``knots". Small-scale,
low-ionization features in planetary nebulae has been considered by
\citet{Balick98}. \citet{Goncalves04} estimated that 10\% of the
Galactic PNe have associated LISs. In an earlier paper,
\citet{Goncalves01} listed 50 PNe occupy different kinds of
small-scale LISs such as jets, tails, filaments, and knots. From the
observation of these LISs, \citet{Goncalves04} summarize their
characteristics as follows:
\begin{enumerate}
\item {There is no preferred distribution for LISs amongst the different PNe morphological classes.}
\item{These structures do not display a density contrast with respect to the main nebular shell, suggesting that both are at the same pressure.}
\item{Most of the LISs studied until 2004 are photoionized.}
\item{Sometimes, they show faster expansion than the main PNe components, but sometimes expand the same velocity.}
\item{LISs appear as pair of jets,  jet-like features, knots, filaments or as isolated systems.}
\end{enumerate}

In the specific case of fast, bipolar or jet-like ejection of LISs,
these were termed fast-moving low-ionisation emission regions or
FLIERs by Balick and co-workers \citep{Balick87, Balick93,
Balick94}, and these have variously been interpreted as jet shocks
in a photoionised medium \citep{Dopita97}, recombination regions in
a mass-loaded jet \citep{Dyson00}, or as ``stagnation knots''
\citep{Steffen02}. However, in other cases, the origin of the knots
is more likely to be found in the late AGB mass ejection phase.
\citet{Miszalski09} found LISs appear to be quite common amongst
post common-envelope PNe and hence they strongly suggest binary
origin of LISs. Recently \citet{Akras16a} showed that the electron
temperatures and chemical abundances of LISs observed in five PNe
are comparable to all nebular components, whereas the electron
density is systematically lower in LISs. They argue that the main
excitation mechanism of LISs is due to shocks, while that of the
other nebular components is due to photoionization.

The main goal of this paper is to study the physical conditions,
chemical composition, kinematical characteristics, ionization
structures, and morphologies of four Galactic PNe: Hen\,2-141 (PN
G325.4-04.0), NGC\,5307 (PN G312.3+10.5), IC\,2553 (PN G285.4-05.3)
and PB\,6 (PN G278.8+04.9), and their associated knots, with the aim
of helping to elucidate the origin of the knots, and their
relationship with the main nebular structures.

The morphologies and structures of these PNe have been previously
described in a number of articles. Narrow-band images of Hen2-141,
in H$\alpha$, [N II] and [O III] emission lines, show a bipolar
morphology with the presence of two symmetrical knots along its
polar axis (Corradi et al. 1996). \citet{Livio97} assign a point
symmetric morphological class for NGC\,5307. The morphology of
IC\,2553 was described as elongated inner shell, surrounded by a
group of knots, with almost a spherical outer shell
\citep{Corradi00}.  Finally,  PB\,6 reveals a relatively circular
double shells in which internal structures appear as a set of knots
\citep{Dufour15}

There have been few previous spectroscopic studies of these PNe, and
almost all rely on the long-silt and \'echelle techniques.
\citet{Costa96} have determined physical conditions and chemical
composition for Hen\,2-141. \citet{Milingo02a, Milingo02b} provide
detailed spectroscopic study for Hen\,2-141 using a long-slit
spectrum covering the range from 3600 to 9600{\AA}. \citet{Ruiz03}
have analyzed the physical conditions and chemical abundances for
NGC\,5307 using \'echelle observations in the spectral range
3100-10360{\AA}. Line fluxes and some nebular parameters of IC\,2553
were determined by \citet{Gutierrez-Moreno85}. Later,
\citet{Perinotto91} analyzed the chemical composition of IC\,2553
and determined its elemental abundances.  \citet{Pena98} have
presented spatially resolved long-slit spectra for PB\,6 to
construct a photo-ionization model for the object. Further, deep
\'echelle spectra in the range 3250-9400{\AA} have been used to
study the physical conditions, chemical composition, and abundance
discrepancy problem in PB6 by \citet{Garcia-Rojas09}. Finally,
\citet{Henry15} have used the HST/STIS spectra to study PB\,6 and
constructing photo-ionization model to predict the mass of its
parent star.

\begin{table}
 \centering
 \small
   \caption{A summary of the observing log.}
    \label{Table1}
   \scalebox{0.85}{
  \begin{tabular}{llccc}
 \hline
   Object  & PNG number & No. of  & Exposure  & Airmass \\
    & & frames &  time (s) &  \\
   \hline
Hen\,2-141 & PN G325.4-04.0 &6 & 300 & 1.13     \\
NGC\,5307  & PN G312.3+10.5 & 6     & 300    & 1.07     \\
IC 2553   & PN G285.4-05.3 & 3     & 600    & 1.12     \\
          &                & 2     & 300    & 1.4      \\
PB\,6      & PN G278.8+04.9 & 3     & 50     & 1.13     \\
          &                & 1     & 300    & 1.14     \\ \hline
 \end{tabular}}
\end{table}

This paper is structured as follows. The observations and data
reduction are given in Section 2. Line fluxes, excitation
properties, discussion of the excitation mechanism of the knots, and
their derived chemical abundances are discussed in Section 3. The
kinematical characteristics are explored in Sections 4. Section 5
provides the morphologies as well as a general discussion for the
results of the four nebulae and their knots. The misclassification
of the CS of IC\,2553 as a weak emission-line star (WELS) is
demonstrated in Section 6, and the conclusions are given in the
final Section 7.

\section{Observations \& data reduction}

The IFU data cubes of the southern planetary nebulae Hen\,2-141,
NGC\,5307, IC\,2553, and PB\,6 were obtained over two nights of 2013
March 28 and 30 using WiFeS instrument mounted on the 2.3 m ANU
telescope at Siding Spring Observatory.   This dual-beam image
slicing integral field spectrograph and its on-telescope performance
are described by \citep{Dopita07, Dopita10}. It provides a 25 arcsec
$\times$ 38 arcsec field of view at spatial resolution of 1.0 arcsec
$\times$ 0.5 arcsec or 1.0 arcsec $\times$ 1.0 arcsec. The observed
data provide low spectral resolution (R $\sim$ 3000) that
corresponds to a full width at half-maximum (FWHM) of $\sim$ 100
km\,s$^{-1}$ ($\sim$ 1.5 {\AA}) for the blue spectral region
3400-5700{\AA}, and high spectral resolution (R $\sim$ 7000) that
corresponds to a FWHM of $\sim$ 45 km s$^{-1}$ ($\sim$ 0.9 {\AA})
for the red spectral region 5500-7000{\AA}. A summary of the
spectroscopic observations is given in Table \ref{Table1}.

\begin{table*}
\centering \caption{Reddening coefficients, observed $H\beta$ and
$H\alpha$ fluxes, and excitation class.} \label{Table2}
\scalebox{0.8}{
\begin{tabular}{lccccccccc}
 \hline
      {\bf Object}        & \multicolumn{2}{c}{$ {c(H\beta)}$} && \multicolumn{2}{c}{ Log F$(H\beta)$}
      && \multicolumn{2}{c}{ Log F$(H\alpha)$} & Excitation class \\

      \cline{2-3} \cline{5-6} \cline{8-9}
           &                &                  &&                              &&       &             &          \\
               & This Paper & Literature       && This paper & Literature  && This paper & Literature & \citet{Reid10}\\
 \hline
Hen\,2-141  &0.89$\pm$0.03  & 0.76$^{(1)}$, 0.57$^{(2)}$ && -11.78 & -11.65$^{(3)}$, -12.11$^{(4)}$ &&-11.05 & -10.99$^{(5)}$   & 11.0\\
NGC\,5307   & 0.55$\pm$0.09 & 0.59$^{(1)}$, 0.42$^{(4)}$ && -11.29 & -11.18$^{(3)}$, -11.61$^{(4)}$ && -10.66 & -10.66$^{(5)}$  & 8.3 \\
IC 2553    & 0.42$\pm$0.08 & 0.49$^{(1)}$, 0.35$^{(6)}$ && -10.88 & -10.82$^{(3)}$            && -10.29 & -10.25$^{(5)}$  & 8.2 \\
PB\,6       & 0.61$\pm$0.04 & 0.52$^{(1)}$, 0.54$^{(7)}$ && -11.93 & -11.87$^{(3)}$, -12.22$^{(8)}$ && -11.28 &  -11.26$^{(5)}$ & 13.6 \\
 \hline
\end{tabular}}
\begin{minipage}[!t]{16cm}
{\small (1) \citet{Tylenda92};(2) \citet{Costa96}; (3)
\citet{Cahn92}; (4) \citet{Milingo02a}; (5) \citet{Frew13}; (6)
\citet{Martin81}; (7) \citet{Kaler91};(8) \citet{Milingo10}\\}
\end{minipage}
\end{table*}

All the data cubes were reduced using the {\tt PYWIFES} data
reduction pipeline \citep{Childress14}. The STIS spectrophotometric
standard stars HD 111980 and HD 074000 were used to calibrate the
flux intensities, while the Cu-Ar arc lamp, with 40s exposures
during the night, was used to calibrate the wavelength scale. The
telluric absorption features arise from atmospheric oxygen and water
vapour molecules were removed from the observations utilizing the
B-type telluric standard HIP 54970 and HIP 66957 with the
spectrophotometric stars as a secondary standard. The final data
cubes were treated from the effects of cosmic rays, sky background,
and instrumental sensitivity in both the spectral and spatial
directions.

\section{Physical and chemical analysis}

\subsection{Line fluxes and excitation properties}

The global spectra of the PNe set were extracted from their specific data cubes using {\tt QfitsView}
software (which is a fits file viewer using the QT widget library developed at the Max Planck
Institute for Extraterrestrial Physics by Thomas Ott). The red spectra were re-scaled, by a
factor $\sim 1-2\%$, to compensate the continuum level of blue spectra using the emission
lines in the common spectral region (5500-5700{\AA}). Emission-line fluxes and their
uncertainties were measured, from the final combined, flux-calibrated blue and red
spectra, using the {\tt  ALFA} code \citep{Wesson16}.  This code uses a group of
emission lines which are expected to be present to construct a synthetic spectrum.
The parameters used to build the synthetic spectrum are developed by a genetic algorithm.
Uncertainties are estimated using the noise structure of the residuals.

The Nebular Empirical Abundance Tool (NEAT; \citet{Wesson12}) was
applied to derive the interstellar reddening coefficients and
subsequent plasma diagnoses. The line intensities were corrected for
reddening applying the extinction law of \citet{Howarth83}. The
reddening coefficient ${c(H\beta)}$ was determined from the ratios
of the Hydrogen Balmer lines (assuming Case B at $T_e = 10^4$\,K),
in an iterative method. In Table \ref{Table2}, we compare the
derived reddening coefficients and H$\alpha$ and H$\beta$ fluxes, on
a log scale, with those given in the literature. In general, our
results agree well with other studies, except slightly higher
reddening coefficients derived for Hen\,2-141 and PB\,6.

In addition to the global spectra of the PNe sample we able to
extract the integrated spectra of six knots,  three belong to
NGC\,5307 (K$_{\textrm{NW}}$, K$_{\textrm{SW}}$, K$_{\textrm{SE}}$),
two belong to IC\,2553 (K$_{\textrm{NE}}$ and K$_{\textrm{SW}}$),
and one associated with PB\,6 (K$_{\textrm{SE}}$). In subsequent
discussions, we will consider only these knots. Results relating to
the knots should be taken with caution because their extracted
spectra may be contaminated by the surrounding nebular gas. A rough
estimate was made to the probable effect of the surrounding nebular
emission to the knots emission that studied in the paper. The
results show the contamination of the surrounding gas to NGC 5307
knots are in the range 10-17\%, in IC 2553 knots 14-19\%, and in PB6
knot 15-18\%. In all knots we see enhancement of the low ionization
line fluxes of ([O I], [O II], [N II], and [S II]) and a diminishing
of the relative He II 4685{\AA} compared with their parent nebulae.
The list of PNe line fluxes and their associated knots are given in
the Tables \ref{TableA1} and \ref{TableA2} in the Appendix.

PNe excitation classes were determined following the \citet{Reid10}
scheme. He II $\lambda$4686/H$\beta$ line ratio probably provides
the best indication for the nebular excitation class (EC). The He II
$\lambda$4686 line disappears completely in the spectra of low
excitation PN (EC $< 5$). \citet{Reid10} remarked that at EC $\geq$
5 [O III]/H$\beta$ line ratio increasing to some degree with nebular
excitation, therefore, they incorporate this ratio with the He II
$\lambda$4686/H$\beta$ line ratio in classifying PNe with EC $\geq$
5. The results show very high excitation classes for PB\,6 and
Hen\,2-141 and  mid to high excitation classes for IC\,2553 and
NGC\,5307 (Table \ref{Table2}).

\subsection{Excitation mechanism of LISs}

\begin{figure*}
    \includegraphics[width=1.0\textwidth]{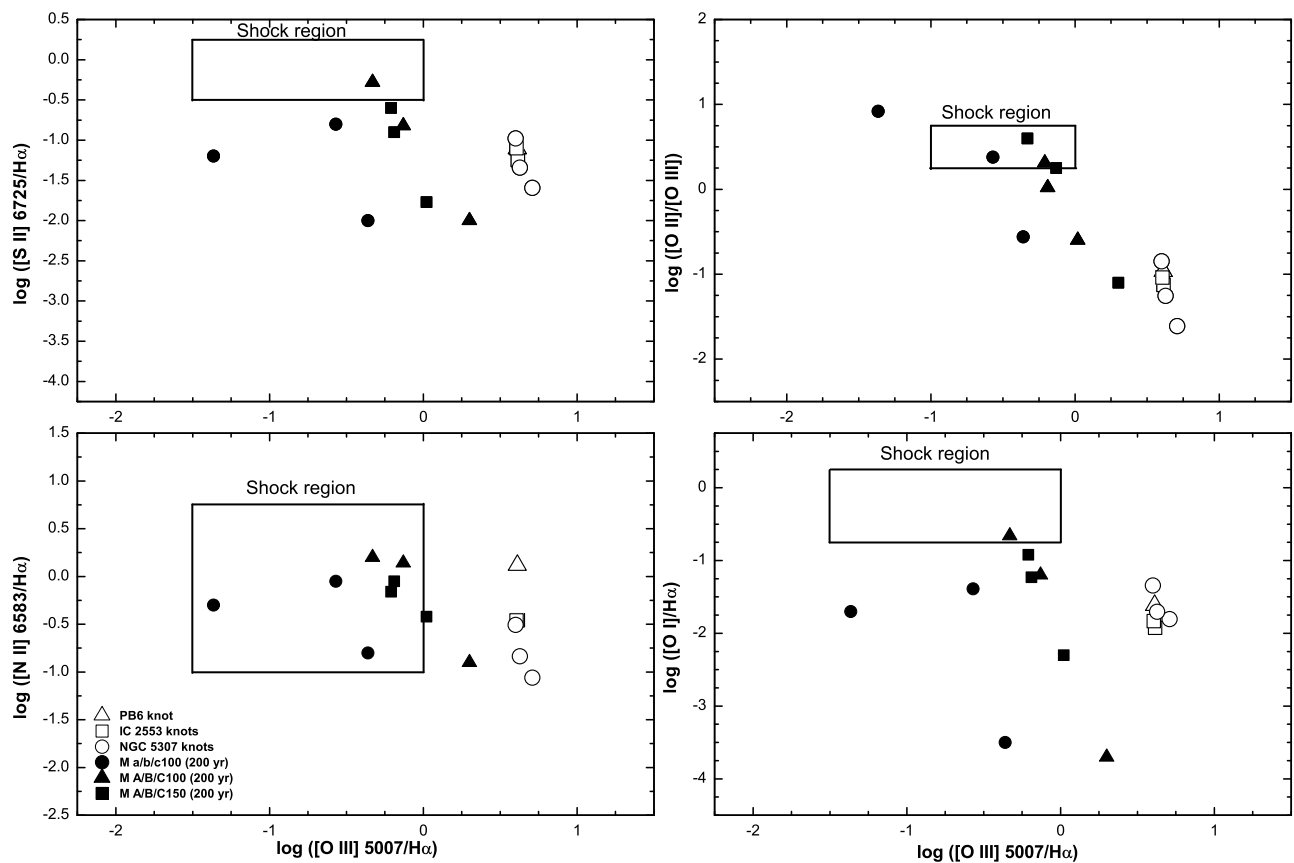}
  \caption{The locations of the knots on the diagnostic diagrams proposed by \citet{Raga08} designed
 to distinguish the shock-excited and photoionized regions. The empty symbols  refer to the knots associated with PB 6 (triangle),
 IC 2553 (squares), and NGC 5307 (circles). The rectangle in each diagram refers to the shock
region. The filled symbols represent the predicted emission line
ratios by \citet{Raga08}; filled circles = models a/b/c/100 (t=200
yr);  filled triangles = models A/B/C/100 (t=200 yr); filled squares
= models A/B/C/150 (t=200 yr). All knots are far away from the
shock-excited regions. This implies that the excitation mechanism
responsible for producing these LISs is more likely to be
photoionisation from the CSs of their nebulae. }
  \label{figure1}
\end{figure*}

In order to examine the excitation mechanisms of the LISs studied
here, we have used the diagnostics technique given by
\citet{Raga08}. The purpose of these diagnostic diagrams is to
discriminate photoionized nebulae from shock-excited regions. Figure
\ref{figure1} shows that the position of all knots are compatible
with them lying in the regime of photoionized plasmas. Thus, all the
knots studied here are likely to be excited by their central stars.
\citet{Raga08} proposed the central star of PN is capable of
producing emission line ratios analogous to those of shocked-excited
nebulae provided that the local ionisation parameter is sufficiently
low. These results are also consistent with the conclusion given by
\citet{Goncalves04} which also reported that most LISs systems are
mainly photoionized. The other common diagnostic diagram produced by
\citet{Sabbadin77} also gives similar results to those given above.
In addition, we found that almost all knots occupy the empirical
zone of fast FLIERS in the \citet{Raga08} diagnostic diagrams.

\begin{table*}
\centering \caption{Temperatures and densities of the sample.}
 \label{Table3}
\scalebox{0.55}{
\begin{tabular}{llllllllllllllllllllllll}
\hline
\\
    & \multicolumn{3}{c}{$N_e$[O~{\sc ii}]} && \multicolumn{3}{c}{$N_e$[S~{\sc ii}]}
    && \multicolumn{3}{c}{$N_e$[Cl~{\sc iii}]} && \multicolumn{3}{c}{$N_e$[Ar~{\sc
    iv}]} && \multicolumn{3}{c}{$T_e$[N~{\sc ii}]} && \multicolumn{3}{c}{$T_e$[O~{\sc
    iii}]} \\

\cline{2-4} \cline{6-8} \cline{10-12} \cline{14-16}
\cline{18-20}\cline{22-24} \\

\hline
Hen\,2-141   &   1932    &   (   1666    )   &   (   -843    )   &&   1355    &   (   160 )   &   (   -143    )   &   &   1184    &   (   2291    )   &   (   -1169   )   &   &   1347    &   (   335 )   &   (   -330    )   &   &   11107   &   (   666 &   (   -622    )   &   &   12160   &   (   92  )   &   (   -92 )   \\

NGC\,5307    &   4422    &   (   915 )   &   (   -681    )   &&   3639    &   (   2144    )   &   (   -1214   )   &   &   1388    &   (   1433    )   &   (   -1036   )   &   &   1449    &   (   285 )   &   (   -285    )   &   &   13316   &   (   1189    &   (   -1090   )   &   &   12332   &   (   396 )   &   (   -384    )   \\

K$_{\textrm{NW}}$   &   3933    &   (   2510    )   &   (   -1326   )   &&   1707    &   (   407 )   &   (   -329    )   &   &   1327    &   (   3241    )   &   (   -1346   )   &   &   1510    &   (   813 )   &   (   -732    )   &   &   13093   &   (   681 &   (   -614    )   &   &   12430   &   (   197 )   &   (   -197    )   \\

K$_{\rm SW}$    &   3213    &   (   632 )   &   (   -528    )   &&   3126    &   (   1133    )   &   (   -767    )   &   &   2181    &   (   1214    )   &   (   -1006   )   &   &   1734    &   (   432 )   &   (   -396    )   &   &   13948   &   (   534 &   (   -514    )   &   &   12302   &   (   380 )   &   (   -369    )   \\

K$_{\rm SE}$    &   3710    &   (   4633    )   &   (   -1712   )   &&   2954    &   (   813 )   &   (   -589    )   &   &   1708    &   (   3445    )   &   (   -1712   )   &   &   1662    &   (   889 )   &   (   -808    )   &   &   12827   &   (   581 &   (   -637    )   &   &   12328   &   (   329 )   &   (   -329    )   \\

IC 2553 &   5493    &   (   3113    )   &   (   -2617   )   &&   3242    &   (   1382    )   &   (   -874    )   &   &   4030    &   (   1397    )   &   (   -1189   )   &   &   4502    &   (   315 )   &   (   -294    )   &   &   11033   &   (   739 &   (   -998    )   &   &   10823   &   (   152 )   &   (   -152    )   \\

K$_{\rm NE}$    &   4233    &   (   2876    )   &   (   -1489   )   &&   3054    &   (   1463    )   &   (   -889    )   &   &   3187    &   (   1296    )   &   (   -1026   )   &   &   3456    &   (   1738    )   &   (   -1428   )   &   &   10719   &   (   598 &   (   -591    )   &   &   10853   &   (   345 )   &   (   -345    )   \\

K$_{\rm SW}$    &   3314    &   (   1946    )   &   (   -1092   )   &&   1744    &   (   781 )   &   (   -539    )   &   &   2262    &   (   960 )   &   (   -782    )   &   &   3410    &   (   1854    )   &   (   -1590   )   &   &   10867   &   (   484 &   (   -484    )   &   &   11012   &   (   344 )   &   (   -344    )   \\

PB\,6    &   2190    &   (   524 )   &   (   -423    )   &&   1765   &   (   244 )   &   (   -215    )   &   &   1932    &   (   1414    )   &   (   -1992   )   &   &   1952    &   (   361 )   &   (   -351    )   &   &   11909   &   (   397 &   (   -397    )   &   &   14778   &   (   151 )   &   (   -151    )   \\

K$_{\rm SE}$    &   1972    &   (   1545    )   &   (   -879    )   &&   1513    &   (   436 )   &   (   -339    )   &   &   1881    &   (   1387    )   &   (   -1158   )   &   &   1642    &   (   1636    )   &   (   -1407   )   &   &   11154   &   (   423 &   (   -423    )   &   &   13751   &   (   541 )   &   (   -541    )   \\

\hline
\end{tabular}}
\end{table*}

\begin{table*}
\centering \caption{The total abundances, in the form of
log(X/H)+12, of the sample and their knots compared with previous
studies.} \label{Table4} \scalebox{0.7}{
\begin{tabular}{lllllllllllllllllllll}
\hline
Element & \multicolumn{3}{c}{Hen\,2-141} && \multicolumn{6}{c}{NGC\,5307} && \multicolumn{4}{c}{IC 2553} && \multicolumn{4}{c}{PB\,6} \\

\cline{2-4} \cline{6-11} \cline{13-16}\\

        &   PN    &   (1)       &   (2)   &&  PN      & K$_{\textrm{NW}}$  &   K$_{\rm SW}$&  K$_{\rm SE}$ & (2)       & (3)       &&  PN      & K$_{\rm NE}$  & K$_{\rm SW}$  &   (4)     &&  PN      & K$_{\rm SE}$  &   (5)     &   (6) \\
\hline
He/H    &   11.04  &  11.11   &   11.08   &&  11.00   &  11.04    &   11.08   &  10.95    &   11.00   &   11.00   &&  11.04   &   11.15   &   11.04   &   11.04   &&  11.23   &   11.18   &   11.26   &   11.23   \\
C/H     &   8.83  &           &           &&  8.55    &   8.28    &   8.68    &   7.81    &           &   8.08    &&  8.91    &   9.03    &   8.77    &   8.91    &&  9.01    &           &   8.92    &           \\
N/H     &   8.19  &   7.98    &   8.41    &&  7.72    &   7.94    &   8.00    &   7.94    &   7.71    &   8.04    &&  8.23    &   8.27    &   8.16    &   8.28    &&  8.58    &   8.54    &   8.62    &   8.67    \\
O/H     &   8.72  &   8.43    &   8.88    &&  8.44    &   8.40    &   8.41    &   8.37    &   8.59    &   8.51    &&  8.59    &   8.79    &   8.54    &   8.80    &&  8.54    &   8.51    &   8.51    &   8.59    \\
Ne/H    &   8.03  &   7.70    &   8.16    &&  7.82    &   8.00    &   7.82    &   7.83    &   7.95    &   7.89    &&  8.01    &   8.34    &   8.09    &   8.14    &&  8.00    &   7.94    &   7.91    &   8.02    \\
Ar/H    &   6.03  &   6.28    &   6.40    &&  5.98    &   5.48    &   6.00    &   6.01    &   6.13    &   5.93    &&  6.39    &   6.55    &   5.76    &           &&  5.95    &   5.96    &   7.59    &   5.97    \\
S/H     &   6.84  &   6.57    &   6.45    &&  6.56    &   6.84    &   6.66    &   6.63    &   6.26    &   7.00    &&  7.00    &   7.07    &   6.92    &           &&  6.88    &   6.71    &   7.11    &   6.64    \\
Cl/H    &   5.25  &           &   5.41    &&  4.44    &   5.26    &   4.54    &   4.90    &   4.83    &   5.00    &&  4.88    &   5.38    &   5.14    &           &&  5.22    &   4.96    &   5.26    &           \\
N/O     &   0.30    &   0.35    &   0.34    &&  0.19    &   0.35    &   0.39    &   0.37    &   0.13    &   0.34    &&  0.44    &   0.31    &   0.41    &   0.30    &&  1.10    &   1.07    &   1.30    &   1.20   \\

\hline
\end{tabular}}
\begin{minipage}[!t]{16cm}
{\small References: (1) \citet{Holovatyy05};(2) \citet{Milingo02b};
(3) \citet{Ruiz03}; (4) \citet{Perinotto91}; (5) \citet{Henry15};
(6) \citet{Perinotto04}; (7) \citet{Grevesse10}\\}
\end{minipage}
\end{table*}

\subsection{Temperatures and densities}

The collisional excitation lines (CELs) and optical recombination
lines (ORLs) identified in the PNe spectra are convenient for their
use in both plasma diagnosis and for elemental abundances
determination. The NEAT code uses the Monte Carlo technique to
propagate the statistical uncertainties of the line fluxes to all
other quantities, e.g. temperatures, densities, and ionic and
elemental abundances. The emission lines detected in each PN
spectrum cover a wide range of ionization states covering neutral
species up to the fourth ionized species, e.g. [O I], [O II], [O
III], [Ne IV] and [Ar V].

The electron temperatures and densities were determined from the
NEAT code. The variety of spectral lines shown in the PNe spectra
permitted us to determine the electron temperatures and densities
from the low and intermediate-ionization zones. The nebular
temperatures were determined from the line ratios [N II]
($\lambda$6548 + $\lambda$6584)/$\lambda$5754 and [O III]
($\lambda$4959 + $\lambda$5007)/$\lambda$4363, while nebular
densities were determined from the line ratios [S II]
$\lambda$6716/$\lambda$6731 and [O II] $\lambda$3727/$\lambda$3729,
[Cl III] $\lambda$ 5517/$\lambda$ 5537, and [Ar IV]
$\lambda$4711/$\lambda$4740. In Table \ref{Table3}, we list the
densities, temperatures, and their uncertainties for the PNe and
their associated knots.

\subsection{Ionic and elemental abundances}

Applying the NEAT code, the ionic abundances of nitrogen, oxygen,
neon, argon, sulfur, and  chlorine can be derived from the CELs,
while helium and carbon were calculated from the ORLs using the
temperature and density relevant to their ionization zone. When
several lines are observed for the same ion the average abundance
was adopted. The total elemental abundances were determined from the
ionic abundances using the ionization correction factors (ICFs)
given by \citet{Delgado-Inglada14}.    The ionic and total
abundances for the PNe as well as knots studied in this paper are
presented in Tables \ref{TableA3} and \ref{TableA4} in the Appendix.

The elemental abundances of the sample, in the form of log(X/H)+12
are presented in Table \ref{Table4} and compared with those found in
the literature. This comparison reveals good agreement with previous
studies. The slight differences remaining can possibly explained due
to the differences in observed line fluxes, as expected between
long-slit and integral field datasets, as well as the different ICFs
and different sources of fundamental atomic data.

The planetary nebula PB\,6 is the only object in our sample is rich
in helium (He/H $\geq$ 0.125) and nitrogen (log(N/H\c)+12 $\geq$
8.0). It also has an N/O ratio $\geq$ 0.5. Therefore it is
classified as Type I according to the original classification scheme
proposed by \citet{Peimbert78}, which probably indicates that the
progenitor star of initial mass $\geq 4 M_{\odot}$. Hen\,2-141 and
IC\,2553 both show an excess in nitrogen abundances (log(N/H)+12
$\geq$ 8.0) and have N/O $\geq$ 0.25. Applying the Peimbert criteria
as modified by \citet{Quireza07} both objects can be classified as
Type IIa. The elemental abundances of NGC\,5307 show this object is
both helium and nitrogen poor. Hence, the object is classified as
being of Type IIb/III.

\begin{table}
\centering \caption{Radial and expansion velocities of the sample.}
 \label{Table5}
\scalebox{0.8}{
\begin{tabular}{lccccc}
 \hline
  Object    & \multicolumn{2}{c}{$RV_{\rm hel}$ (km s$^{-1}$)}& &
 \multicolumn{2}{c}{V$_{exp}$ (km s$^{-1}$)} \\ \cline{2-3} \cline{5-6}
  \\
            &    Ours    &   Literature      && Ours  &  Literature \\
\hline
Hen\,2-141   & -39$\pm$5 & $-46\pm9.0^1$  && 28.0  &  \\
\\
NGC\,5307    & 38$\pm$3  & $39\pm2^1$    && 22.1  &  \\
K$_{\textrm{NW}}$ & 33$\pm$6  &   && &  \\
K$_{\textrm{SW}}$ & 24$\pm$6  &   && &  \\
K$_{\textrm{SE}}$ & 12$\pm$6  &   && &  \\
\\
IC 2553     & 27$\pm$3  & $37\pm6^1$, $31\pm4^2$    && 19.3  &  \\
K$_{\textrm{NE}}$ & 23$\pm$6  &   && &  \\
K$_{\textrm{SW}}$ & 17$\pm$5  &   && &  \\
\\
PB\,6        & 51$\pm$4  & $59\pm1^1$   && 28.7  & $34^3$, $38\pm4^4$  \\
K$_{\textrm{SE}}$ & 60$\pm$6  &   && &  \\
\hline
\end{tabular}}
\begin{minipage}[!t]{8cm}
{\small $^1$\citet{Durand98}; $^2$\citet{Corradi00};
$^3$\citet{Gesicki03}; $^4$\citet{Garcia-Rojas09}\\}
\end{minipage}
\end{table}

\section{Kinematical characteristics}

The systemic velocities RV$_{\rm sys}$ of the sample were determined
using the IRAF external package RVSAO (emsao task). Weighted mean
systemic velocities were calculated from the seven emission lines:
H$\alpha$, [NII] $\lambda$6548, [NII] $\lambda$6583, [S II]
$\lambda$6716, [S II] $\lambda$6731, He I $\lambda$6678, and [O I]
$\lambda$6300. These lines lie in the red spectral region of high
spectral resolution. The Heliocentric radial velocities RV$_{\rm
Hel}$, were calculated by correcting the RV$_{\rm sys}$ for the
effect of Earth's motion, using the heliocentric correction factor
given in the fits image header. In Table \ref{Table5}, we present
our results compared with that of literature. The measurements of
Hen\,2-141 and NGC\,5307 show good agreement with \citet{Durand98}
within the error range, while IC\,2553 and PB\,6 are slightly lower.
The RV$_{\rm Hel}$ of IC\,2553 shows better agreement with the value
derived by \citet{Corradi00}. The small differences appear in case
of IC\,2553 and PB\,6 compared to the values of \citet{Durand98} may
be attributed to the different spectral resolutions as well as the
fact that our measurements were extracted from integrated spectra.
We also determined the RV$_{\rm Hel}$ for all knots. In general, the
knots show lower velocities compared to their associated nebulae,
except the knot SE knot of PB\,6 which has a higher velocity. The
RV$_{\rm Hel}$ of the SW and SE knots of NGC\,5307 have much lower
velocity relative to the entire nebula. These results tend to
suggest that the knots are best seen when they lie on the far side
of the expanding nebula (and therefore seen with a redshift), which
is consistent with them being dense partially-ionised structures
photoionised  on the side nearest the central star.

The expansion velocity is a key parameter in studying the PN
evolution. The average expansion velocities of the sample were
measured from the following emission lines: H$\alpha$, [NII]
$\lambda$6548,  [NII] $\lambda$6583,  [S II] $\lambda$6716,  [S II]
$\lambda$6731 and He I $\lambda$6678 following \citet{Gieseking86}.
Results are given in Table \ref{Table5}. Two measurements were also
found in literature  regarding the average expansion velocity of
PB\,6. Both are slightly larger than our measurement.

\section{Morphologies}
All objects in our sample reveal prominent low-ionization regions in
their nebular shells. To study these features and the PNe
morphologies in general, we extract emission-line maps in species of
differing ionisation potential from their WiFeS data cubes using the
QfitsView software.  In general, the ionization strength decreases
outward from the CS due to the dilution and absorption of the
stellar ionising radiation field. Thus, to reveal the ionization
structures of these PNe, we build a composite color image for every
PN by combining three lines of different ionization potential into
one RGB image.

\subsection{Hen\,2-141}

\begin{figure*}
  \begin{tabular}{@{}cccc@{}}
    \includegraphics[scale=0.5]{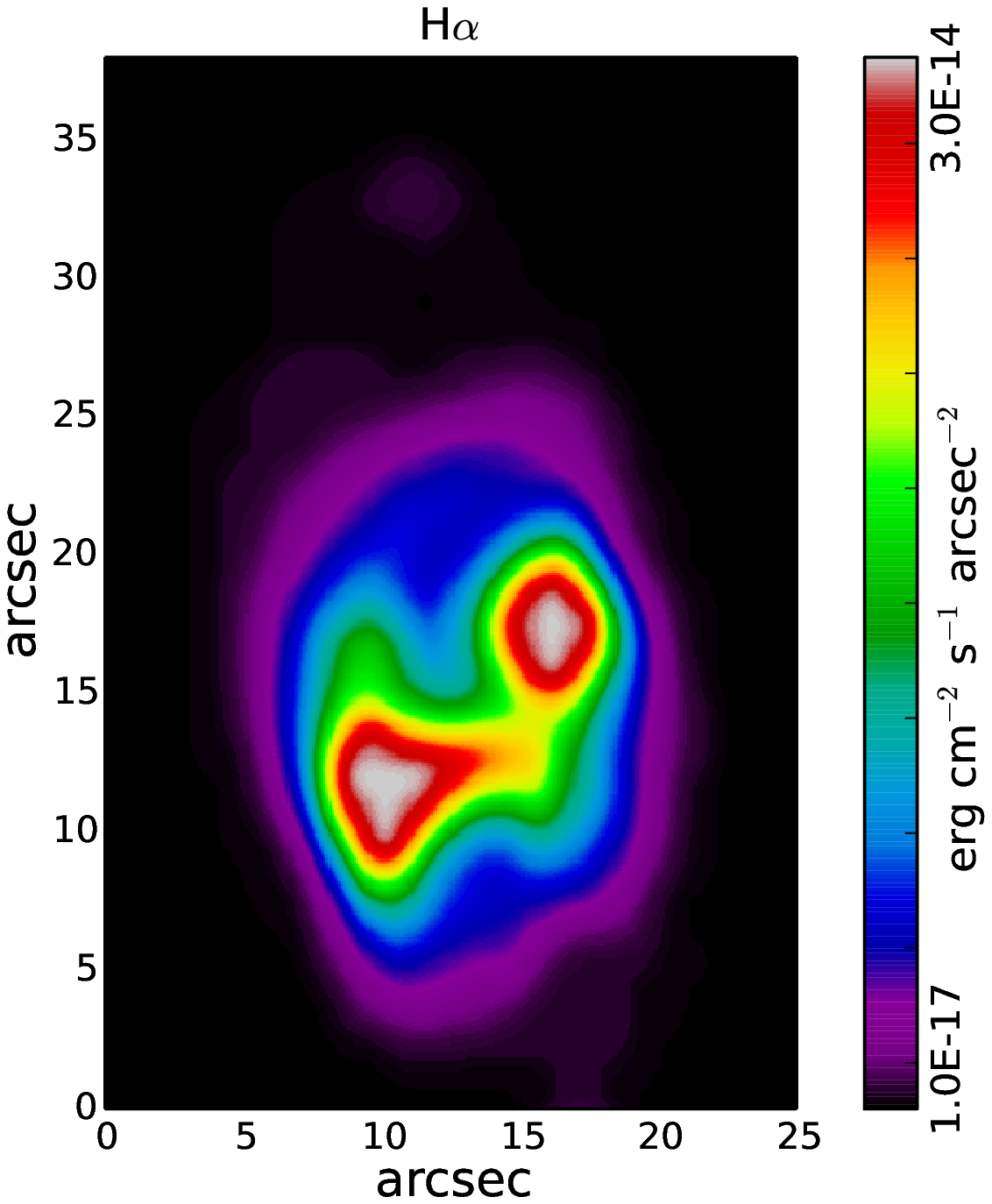} &
    \includegraphics[scale=0.5]{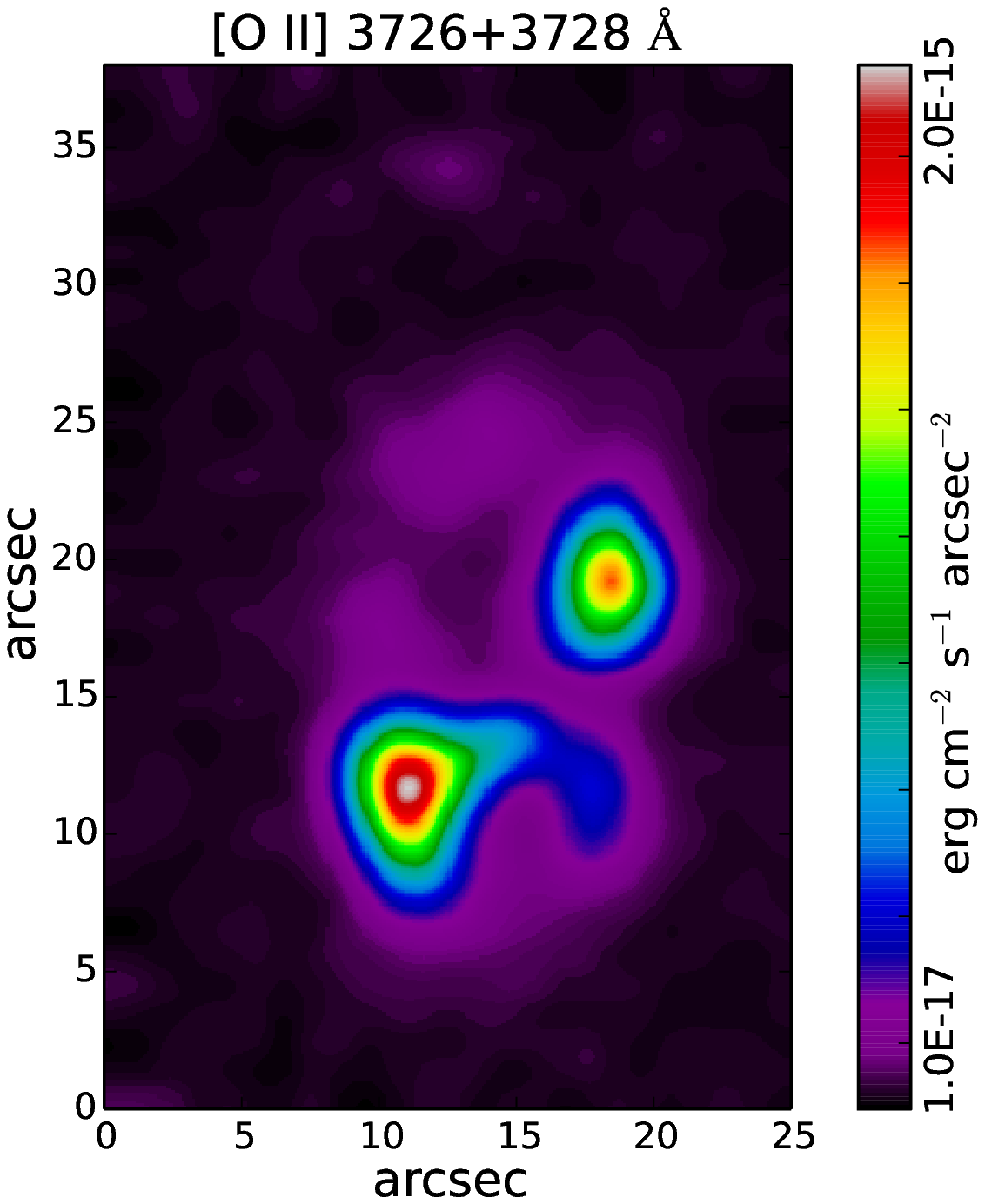} &
    \includegraphics[scale=0.5]{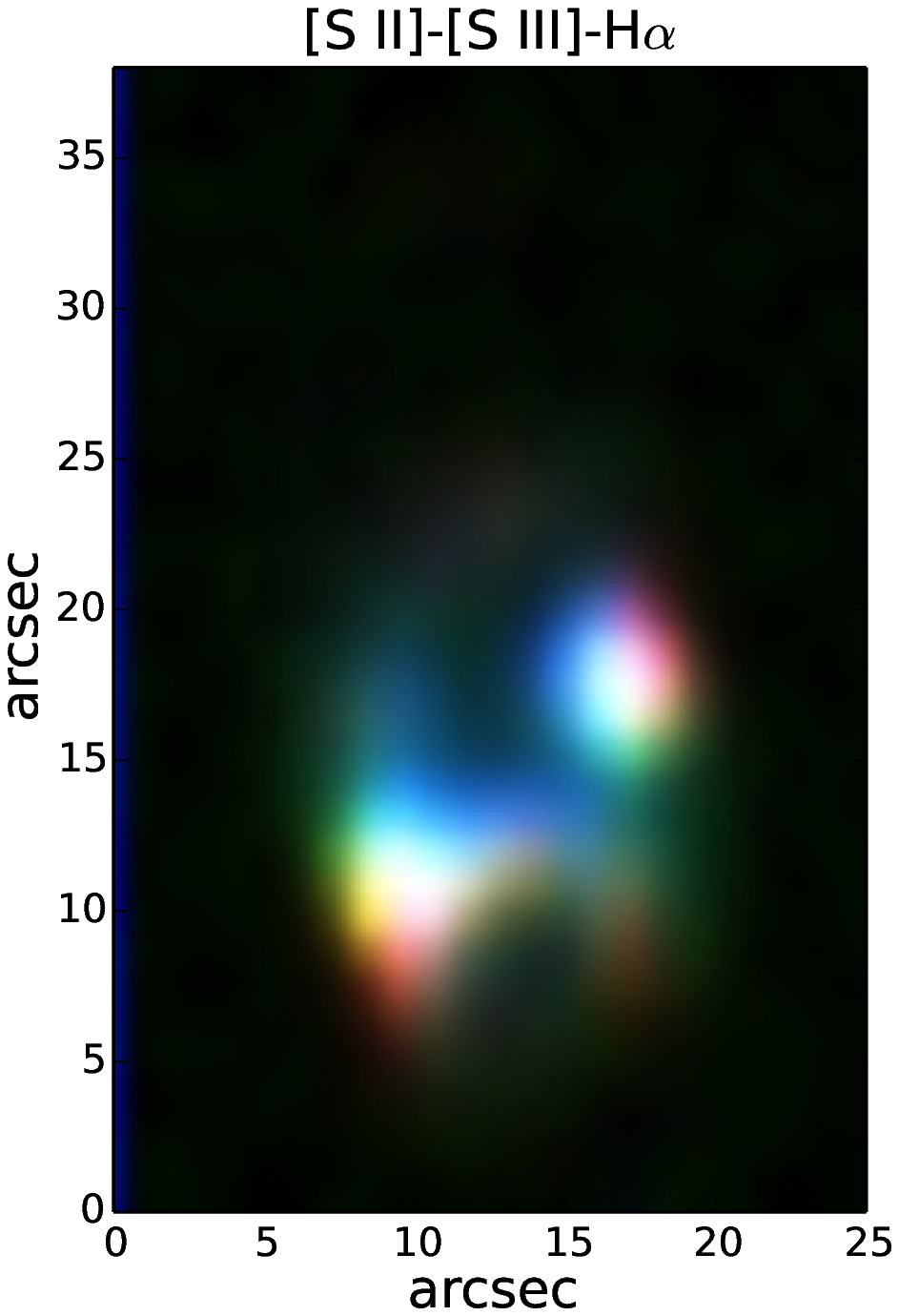}
\end{tabular}
  \caption{Two emission-line maps in different ionization states and composite
 color image for Hen\,2-141. The left and middle panels show the maps
of Hen\,2-141 in H$\alpha$ and [O II] 3726+3728{\AA} lines,
respectively. {\bf The color bar in each figure refers to the
intensity of emission line through the entire nebula. The top and
bottom labels on the color bar represent the maximum and minimum
emission values and the scale is linear.} The right panel shows a [S
II]-[S III]-H$\alpha$ RGB color image of the object. In this and
subsequent figures, north is up and east is on the left. This figure
show a bipolar morphology for the object with two knots to the north
and to the south which barely appear in magenta (R+B) color (right
panel) at the outer limb of the two lobes. The low-ionisation knots
are bi-symmetric with respect to the nebular centre.}
\label{figure2}
\end{figure*}

\citet{Corradi96} have detected two radially symmetrical knots along
the nebular polar axis and two symmetrical ansae along the major
axis of Hen\,2-141. This pair of ansae lies outside the WiFeS field
of view. In Figure \ref{figure2}, we present two emission-line maps
of Hen\,2-141 in the lines of H$\alpha$ (left panel) and [O II]
3726+3728{\AA} (middle panel). The color bar on the right side of
each map refers to the relative surface brightness of emission line
depicted. The PN images in this figure and subsequent figures are
oriented with north up and east to the left. The overall morphology
of the object reveals a bipolar shape where the two lobes appear as
two condensations of gas on the opposite sides of the nebular
center, seen best in [O II] at the top and bottom of the image.
There is also a pair of lower excitation knots at  P.A. $\sim
134^{\rm o}$ associated with the boundary of the inner elliptical
shell.

In Figure \ref{figure2} (right panel), we present a composite color
image of the nebula in the RGB color system ([S II] (red channel),
[S III] (green channel) and H$\alpha$ (Blue channel). The purpose of
this image is to probe the pair of knots and the overall ionization
structures of the nebula. It is obvious the hydrogen gas (blue) is
distributed throughout the PN and extends to the outer region. The
two knots barely appear in magenta (R+B) color at the outer edges of
the two lobes. The white (R+G+B) color of the two lobes are due to
the combination of neutral hydrogen with singly and doubly ionized
sulfur gases. The lowest excitation regions appear red in this
image.

The elemental abundances of the nebula agree well with the results
of \citet{Holovatyy05} and \citet{Milingo02b}, except the argon and
sulfur elements which show slightly lower and higher abundances,
respectively (Table \ref{Table4}). The determined N/O ratio are
consistent with other works. The calculated radial velocity of the
object agrees well with the value of \citet{Durand98}.

\subsection{NGC\,5307}

NGC\,5307 nebula has been classified as belonging to the point
symmetric class: ``the morphological components show point
reflection symmetry about the center"  \citep{Livio97}, based on the
HST image taken by \citet{Bond95}. The same morphology was assigned
by \citet{Gorny97}.

\begin{figure*}
  \begin{tabular}{@{}ccc@{}}
    \includegraphics[scale=0.5]{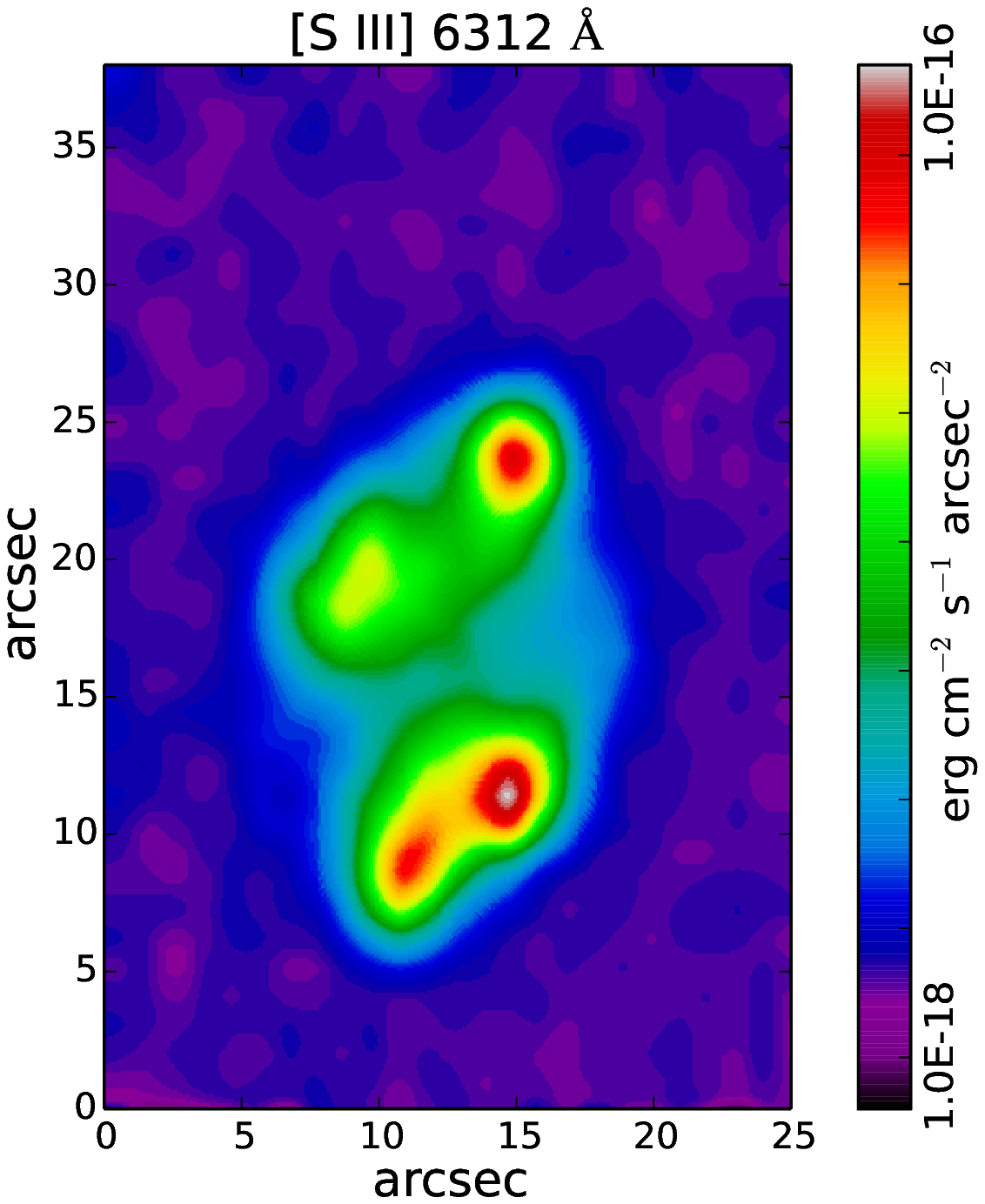} &
    \includegraphics[scale=0.5]{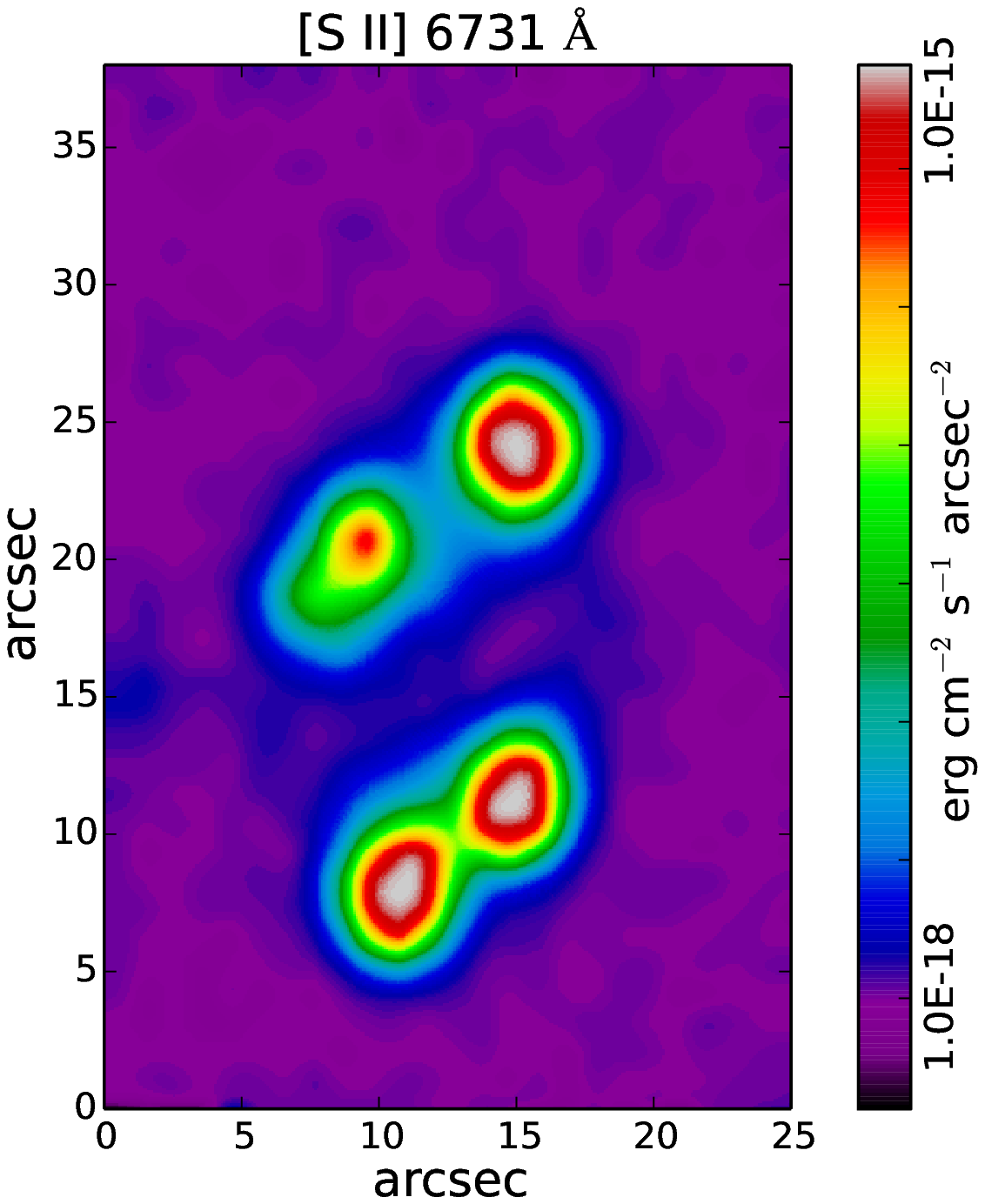} &
    \includegraphics[scale=0.5]{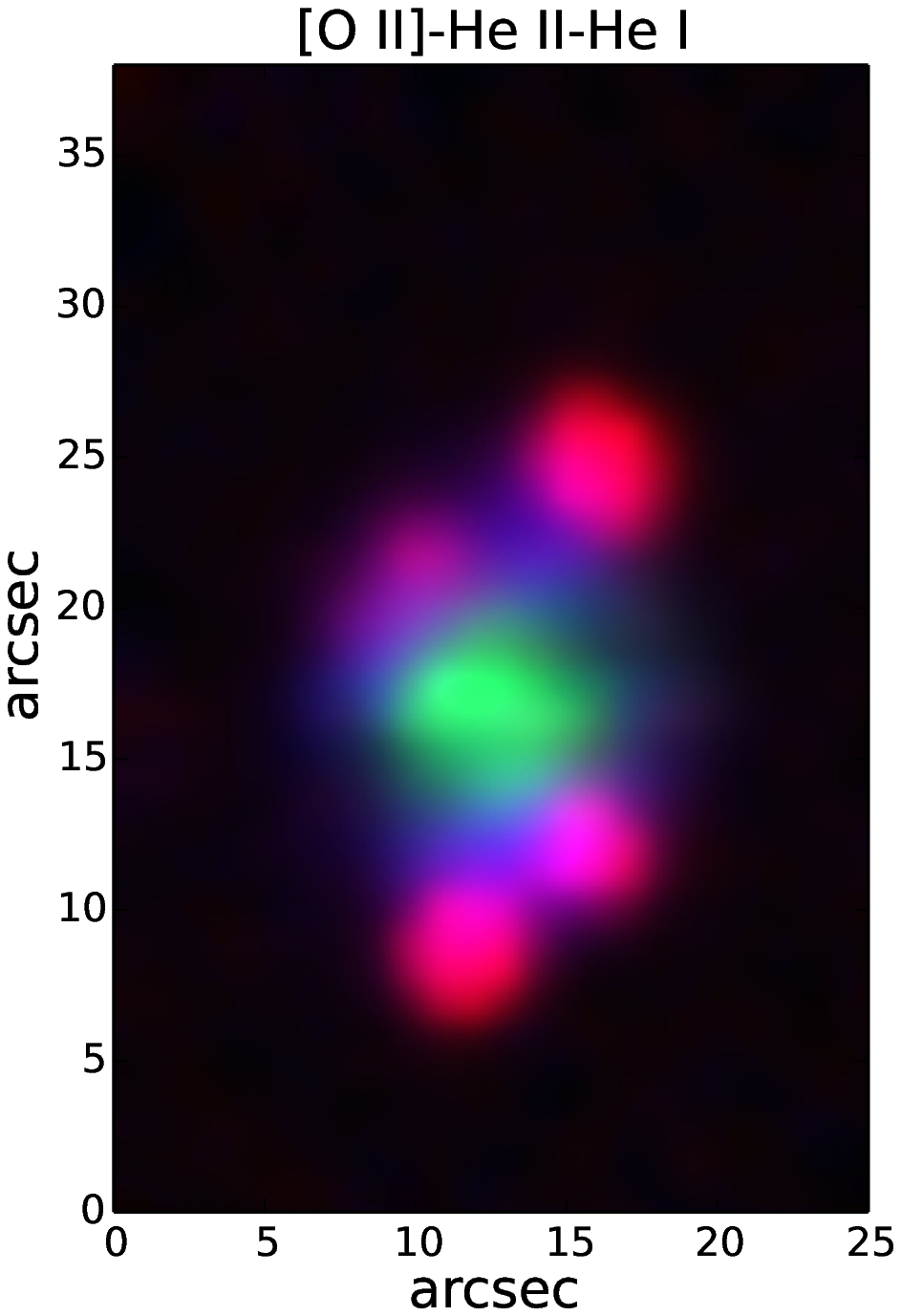}
\end{tabular}
  \caption{As Fig. 2, but for  NGC\,5307. The left and middle panels show the
 maps of NGC\,5307 in [S III] and [S II] lines, respectively. The
right panel shows {\bf an [O II]-He II-He I} RGB color image of the
object. This object shows a bi-symmetric shape with two pairs of
knots on either sides of the nebular center. These four knots are
very pronounced in magenta color ([O II] + He I) in the composite
RGB image (right panel).} \label{figure3}
\end{figure*}

The [S III] and [S II]  emission-line maps (Figure \ref{figure3},
left and right panels) of NGC\,5307 reveal an overall rectangular
shape with two symmetric pairs of knots on either side of the
nebular center at P.A. $\sim 30^{\rm o}$ and at P.A. $\sim 163^{\rm
o}$. The two pairs of knots which appear in Figure \ref{figure3} are
clearly visible in the HST image (Figure 6, \citet{Livio97}).

The four knots are very clear in magenta color of the composite RGB
image (Figure \ref{figure3}, right panel). The NW and SE pair of
knots are much brighter than the NE and SW pair of knots. The
ionization stratification is evident in this image where the
high-ionization helium gas (green color) is distributed in the
center of the nebula, the neutral helium (blue color) lies
throughout the nebula except the inner region, and the
low-ionization oxygen (red color) is confined to the outer region of
the nebula.

The plasma diagnostics given in Table \ref{Table3} show consistency
between the temperature of NW, SW, and SE knots and the nebula as a
whole. However, the electron densities, $N_e$[O~{\sc ii}] and
$N_e$[S~{\sc ii}], of the knots are somewhat lower than the nebula.
Furthermore, the spectroscopy of the knots suggest an enhancement in
the nitrogen abundances compared to the nebula. The elemental
abundances of NGC\,5307 are compared with the results of
\citet{Milingo02b} and \citet{Ruiz03} in Table \ref{Table4}. It is
appropriate to mention here that the nebular spectra observed by
those authors are much deeper than our IFU spectra, and cover the
optical and near IR spectral range from 3100 to 10360\AA. Our
derived abundances of almost all elements nearly match that derived
by those authors, except for  Cl, which has lower abundances (Table
\ref{Table4}). The derived N/O ratio is higher than
\citet{Milingo02b} but lower than \citet{Ruiz03}. The radial
velocity determined from IFU spectra is fully consistent with that
of \citet{Durand98}.

\subsection{IC 2553}

A spatio-kinematic model was constructed for IC\,2553 by
\citet{Corradi00}. It reveals an elongated inner shell and a roughly
spherical outer shell expanding with higher velocity relative to the
inner one. They provide two narrow band images for IC\,2553. The [O
III] image delineates a roughly rectangular inner shell with two
bulges along the short axis encircled by a faint outer shell, while
the [N II] image shows the inner shell is surrounded by few low
ionization knots, a pair of which are located symmetrically to the
CS.

\begin{figure*}
  \begin{tabular}{@{}ccc@{}}
    \includegraphics[scale=0.5]{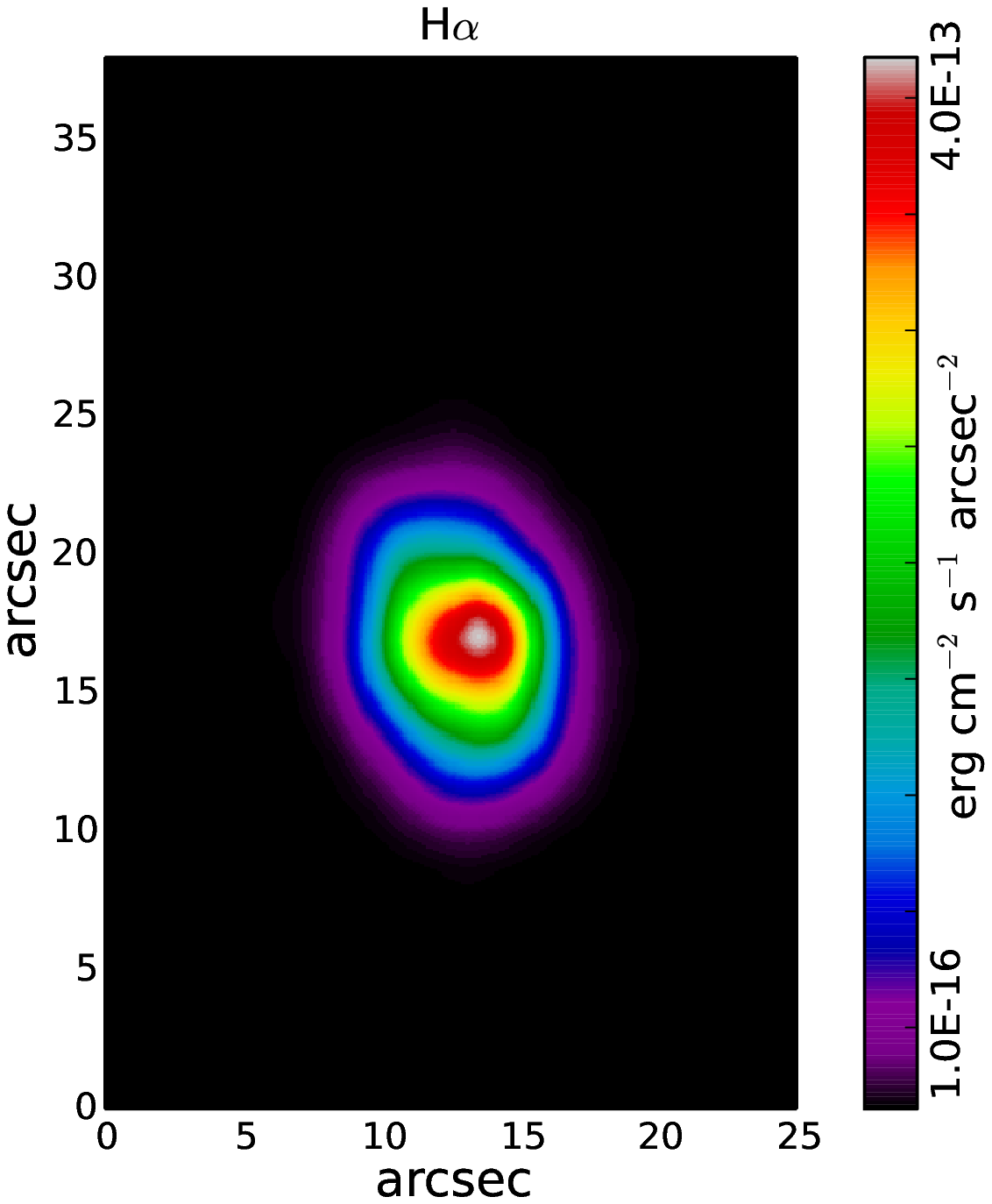} &
    \includegraphics[scale=0.5]{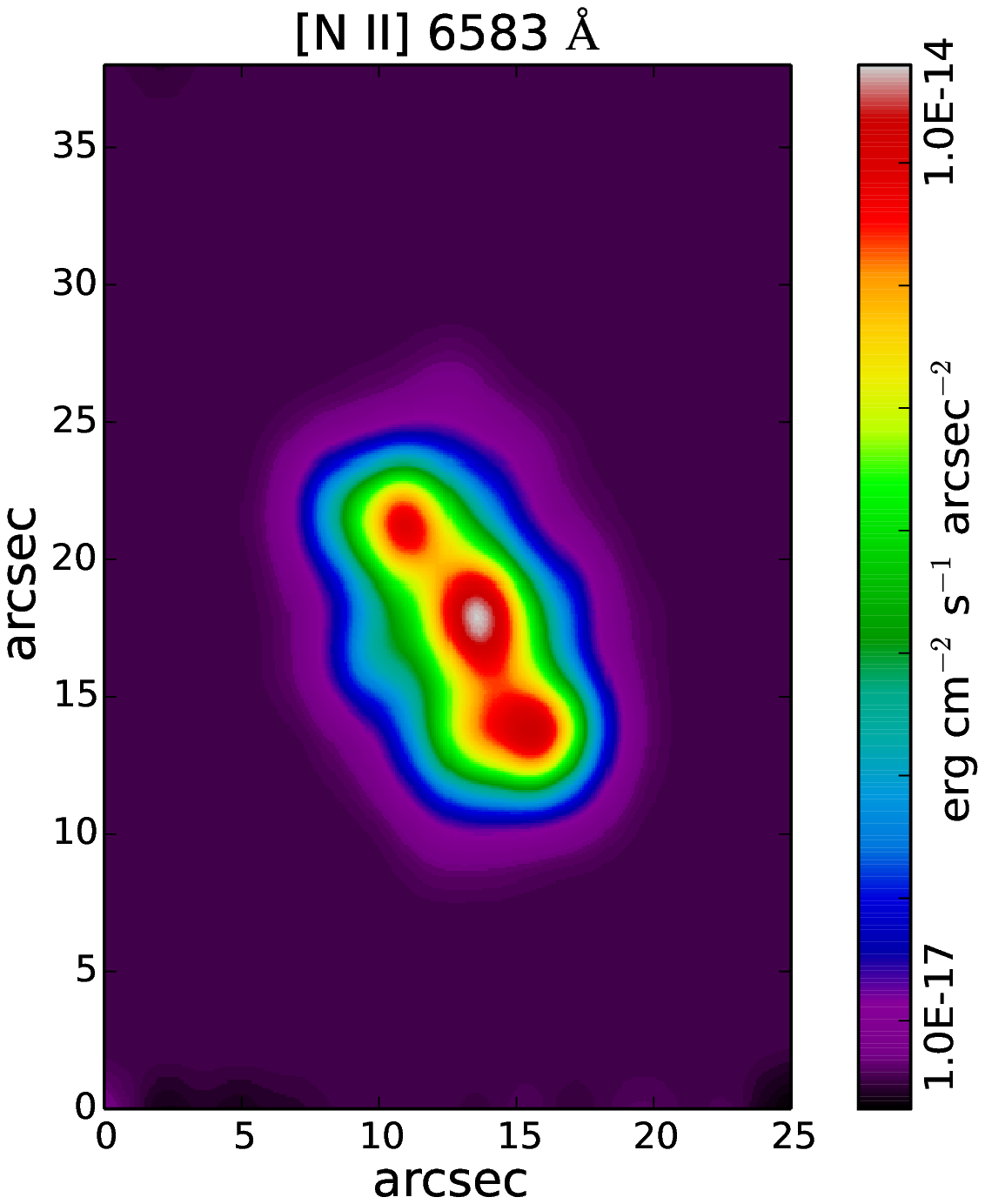} &
    \includegraphics[scale=0.5]{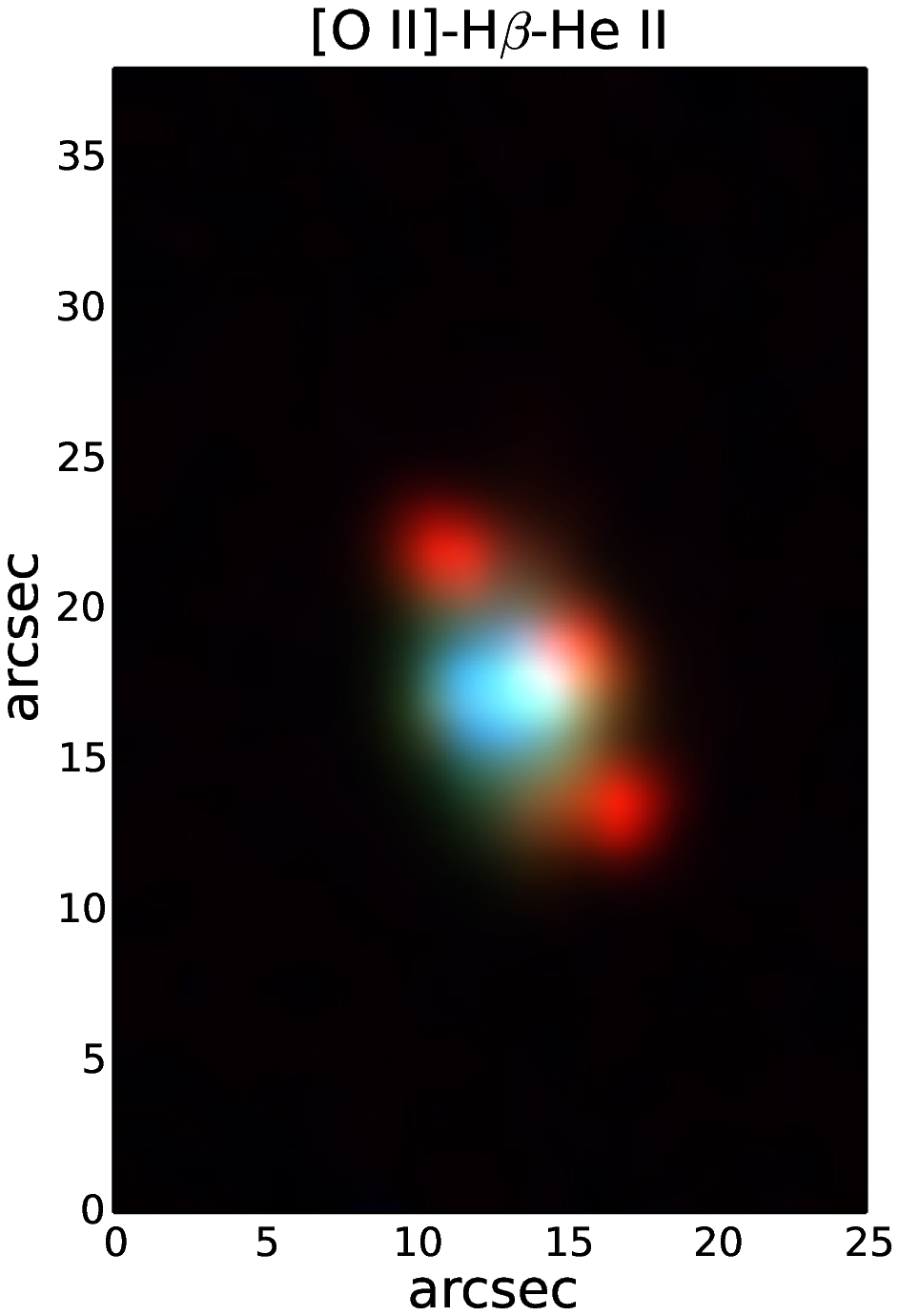}
\end{tabular}
  \caption{As Fig. 2, but for  IC\,2553. The left and middle panels show the
maps of IC2553 in the H$\alpha$ and [N II] 3726+3728{\AA} lines,
respectively. The right panel shows a [O II]-H$\beta$-He II RGB
composite color image of the object. The object show an elliptical
morphology with two isolated knots (in red color - right panel) lie
symmetrically along the major axis. Further, there is another
remarkable low-excitation knot (seen in red)  on the outer limb of
the central spherical region in the north-west direction.}
\label{figure4}
\end{figure*}

Our emission-line maps of IC\,2553 in H$\alpha$ and [N II] 6583
{\AA} lines were given in Figure \ref{figure4}. Both images reveal
an overall elliptical morphology. The [N II]  emission is more
extended than H$\alpha$ map with two protruberences at both sides of
the minor axis. Further, this image revealed the presence of two
knots which lie symmetrically along the major axis of the object at
P.A.$\sim 32^{\rm o}$.  In the right panel of Figure \ref{figure4},
we present a composite RGB color image ([O II] (red channel),
H$\beta$ (green channel) and He II (blue channel). The two
symmetrical knots are more evident here than in the [N II] map. Both
knots (in red colour) appear to be nearly isolated from the central
regiont.We also note the presence of a faint low ionization
structure located on the outer edge of the spherical region in the
north-westerly direction.

The temperatures of NE and SW knots are both compatible with the
entire nebula, while their densities are lower than the nebula. In
general, the elemental abundances of IC\,2553 agree well with the
results of \citet{Perinotto91}, except for the N/O ratio which
appears much higher here. This is consistent with the fact that we
include all of the [N II] - emitting zone in our observations.
Furthermore, the abundances of the two knots are consistent with the
nebula as a whole. As noted above, the radial velocities of both
knots are smaller than the main shell.

\subsection{PB\,6}

\begin{figure*}
  \begin{tabular}{@{}ccc@{}}
    \includegraphics[scale=0.5]{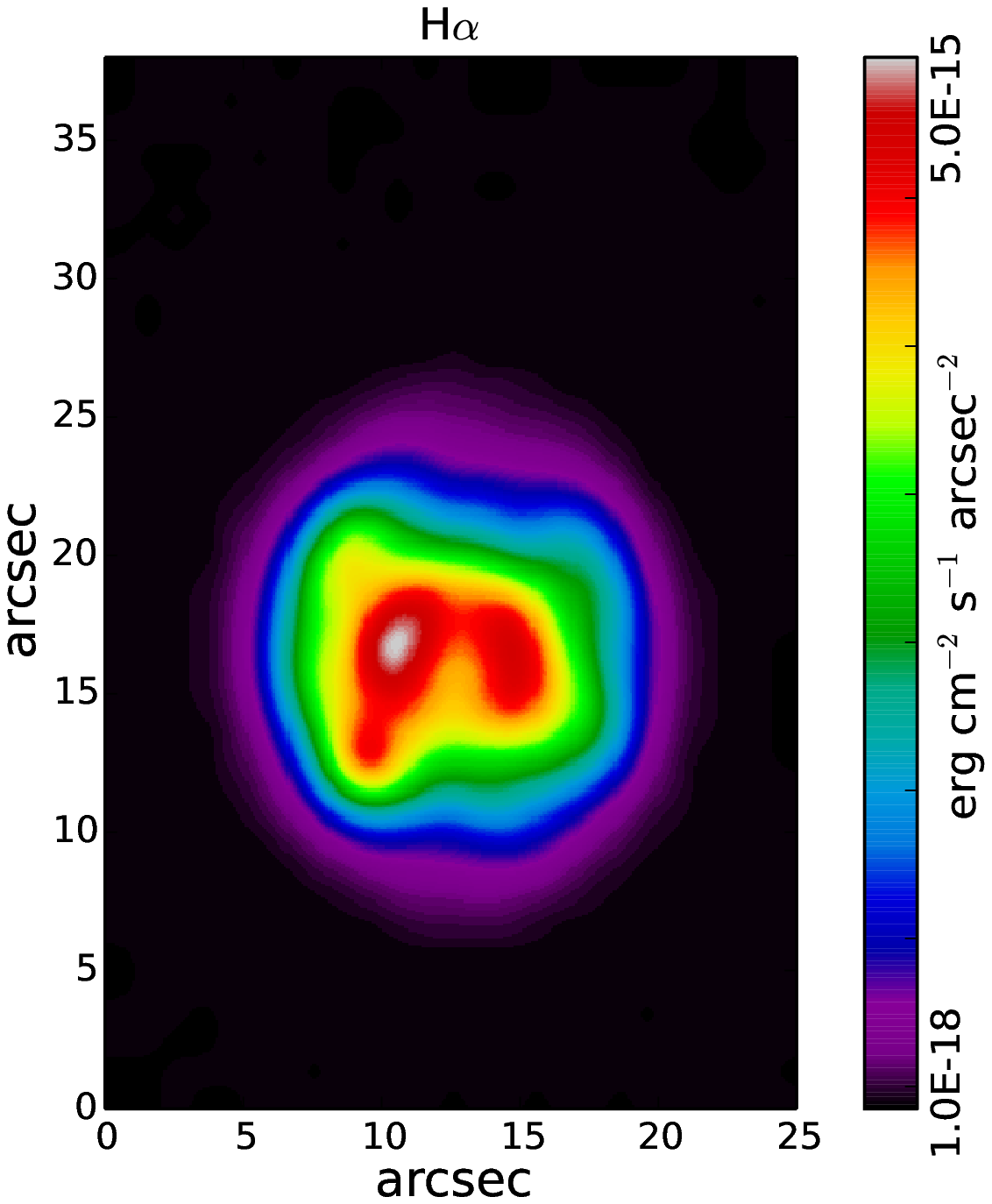} &
    \includegraphics[scale=0.5]{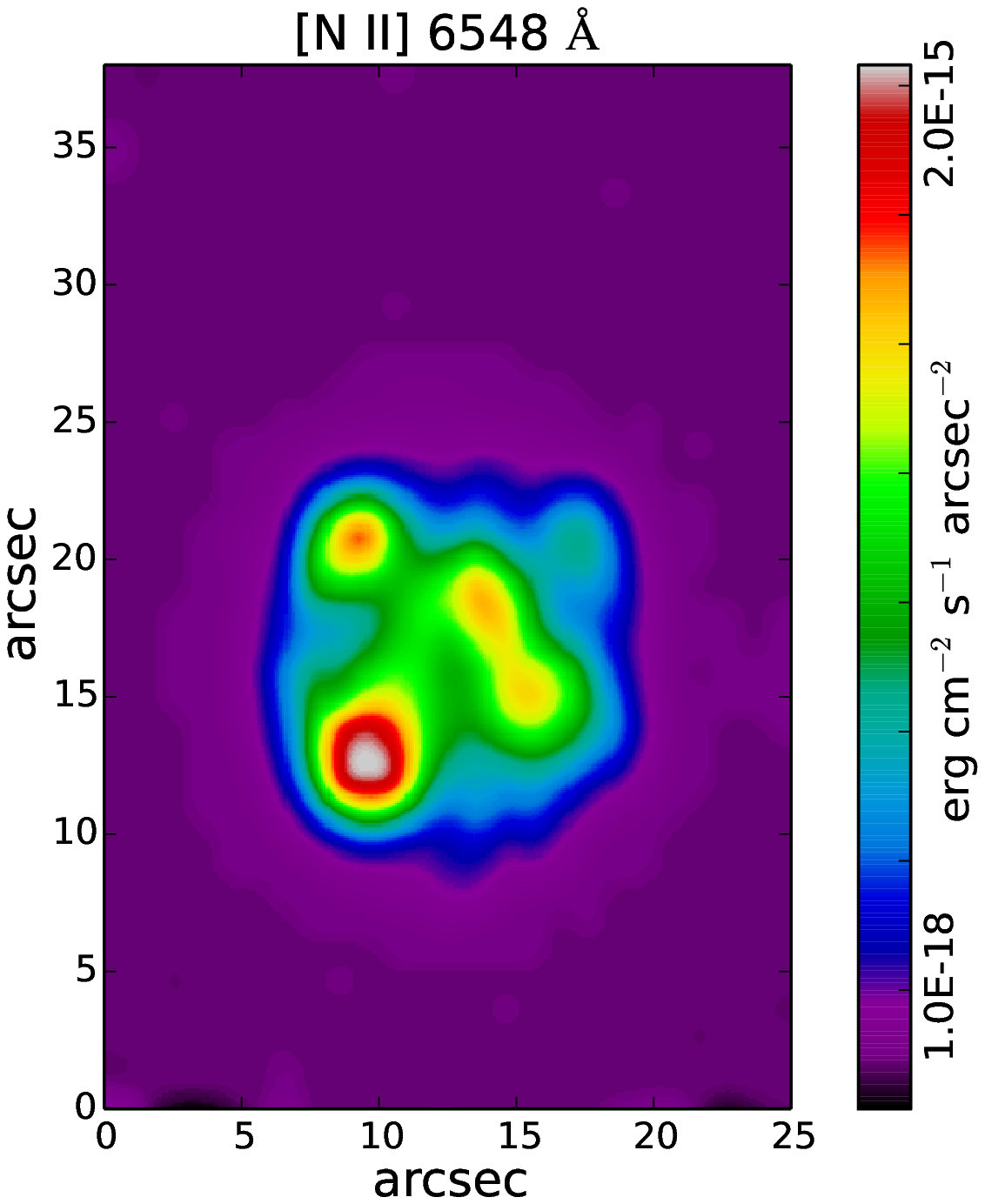} &
    \includegraphics[scale=0.5]{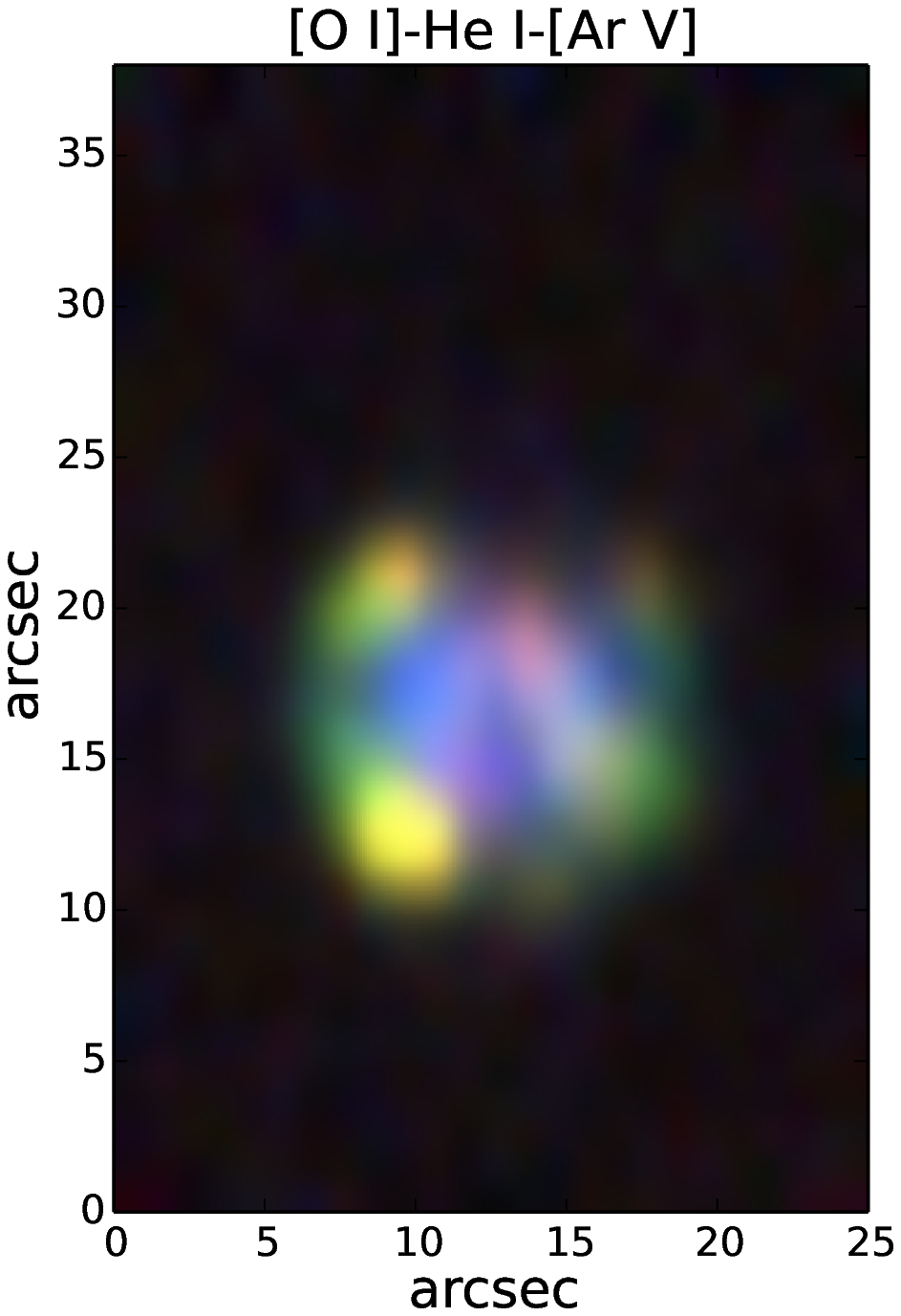}
\end{tabular}
  \caption{As Fig. 2, but for  PB\,6. The left and middle panels show the maps of the
nebula in H$\alpha$ and [N II] lines, respectively. The right panel
shows {\bf an [O I] - He I - [Ar V]} RGB color image of the object.
A roughly circular PN morphology with faint extended halo (appears
in [N II] maps) and three non symmetric knots appear inside the
nebula. Two bright knots are pronounced in yellow (R+G) color and
the third knot appears in magenta (R+B) color with a tail of magenta
and green colors} \label{figure5}
\end{figure*}

The ground-based narrow-band image of PB\,6 in [O III] filter
\citep{Dufour15}, shows two concentric nearly circular shells. These
authors also present a higher-resolution STIS/HST image which shows
a complex system of knots inside the nebula. Figure \ref{figure5}
shows the morphology of PB\,6 as a roughly circular PN with faint
extended halo appearing in the [N II] emission-line map (middle
panel). This map reveals also three non symmetric knots inside the
main ionized shell of the nebula. One of the knots has a tail-like
extension. The [O I]-He I- [Ar V] composite RGB color image (right
panel) shows the same features seen in the [N II] image. The [Ar V]
emission (blue color) fills the central region of the PN, while the
He I emission (green color) is distributed at the outer region and
partially extends into the inner region. Two bright knots are
pronounced on the E side of the nebula in yellow (R+G) color. These
seem to have fainter counterparts on the W side. In addition,
another knot appears N of the nebular centre in magenta (R+B) color
with a tail of magenta and green colors.

The expansion velocity of the entire nebula are slightly smaller
than the values in literature (Table \ref{Table5}). The temperatures
of the SE knot are slightly lower than the nebula as a whole. The
densities of SE knot are marginally smaller than the nebula (Table
\ref{Table3}). The total abundances of PB\,6 are fully consistent
with SE knot as well as the previous studies (\citet{Henry15} and
\citet{Perinotto04}).

\section{Misclassification of IC\,2553 central star}

\citet{Miszalski09} and  \citet{Miszalski11} claim that many of the
weak-emission line CSs type are probably misclassified close
binaries and their characteristic lines (C II at 4267{\AA}, NIII at
4634{\AA} and 4641{\AA}, CIII at 4650{\AA}, CIV at 5801{\AA} and
5811{\AA}) originate from the irradiated zone on the side of the
companion facing the primary. The first evidence for the nebular
origin of most key lines characterize the WELS type was given by
\citet{Gorny14} when he found these lines appear in a spatially
extended region in the 2D spectra of NGC 5979. \citet{Basurah16}
declared that, for many objects, the WELS type may well be spurious.
From WiFeS data of NGC 5979, M4-2 and My60, they showed that the
characteristic CS recombination lines of WELS type are of nebular
origin. They were re-classified the CSs of these nebulae as hydrogen
rich O(H)-Type . \citet{Ali16} found a further example (M3-6) that
further indicates the unreliability of the WELS classification. From
a careful nebular subtracted spectrum of the CS in M3-6, they
revised its classification as H-rich star of spectral type O3 I(f*).
In addition, they show the emission of almost all proposed CS
recombination lines  as WELS type are spatially distributed over a
large nebular area. Here, we present yet another example of
misclassification of WELS type which raises the probability that
most genuine cases of WELS originate from irradiation effects in
close binary central stars.

The CS of IC\,2553 has been classified as of the WELS type by
\citet{Weidmann11}. However, in Figure \ref{figure6}, we present
emission-line maps of IC\,2553 in four CELs (N III 4631+4641{\AA},
CIII 4650{\AA}, O III+O V 5592{\AA}, and C IV 5811{\AA}) which are
supposed to be of CS origin according to the definition of the WELS
type. It is clear that the emission in all of these lines are
spatially distributed over a large area of the PN, except for the C
IV 5811{\AA} line which may truly be of CS origin. Unfortunately we
were unable to extract the CS spectrum from the available data cube
of this PN, but it is clear {\bf that} the WELS type classification
is erroneous for this nebula.

\section{Conclusions}

We have studied the four southern PNe Hen\,2-141, NGC\,5307,
IC\,2553, and PB\,6 using the integral field unit spectroscopy
technique.  We performed emission-line maps as well as composite
color images to study the morphologies and ionization structures of
these objects.  The maps reveal the presence of low-ionization knots
in all objects. Two knots appear in Hen\,2-141, four in NGC\,5307,
two in IC\,2553, and three in PB6.
Furthermore, for first time, we shed the light to the spectroscopy
of six knots associated with NGC\,5307, IC\,2553, and PB\,6. The
derived physical and kinematical properties of these knots should
taken with caution due to the probable contamination of these knots
with their surrounding nebular gases. However, in general, the
physical conditions and elemental abundances of the knots are agree
with their accompanying nebulae, except the electron densities in
the knots are systematically lower. Furthermore we noticed a
slightly higher nitrogen abundances in the knots of NGC\,5307
compared to the entire nebula. Those knots also have lower radial
velocities than their parent nebulae, suggesting that they are dense
part-ionised structures lying on the far side of the expanding
nebula (and therefore seen with a redshift), and are have ionisation
fronts photoionised on the side which lies nearest the central star.

The physical analysis of the entire nebulae indicate a  medium to
high excitation class for NGC\,5307 and IC\,2553 and very high
excitation classes for Hen\,2-141 and PB6. Their elemental
abundances compare relatively well with those previously published
in literature.

Another example for the misclassification of WELS type central stars
was given here. The demonstrated emission-line maps of IC\,2553 in
the collisional characteristic lines of WELS type show that these
spectral lines are spatially distributed over a wide nebular area
and therefore are not of CS origin but arise from within the nebula.

\section*{acknowledgements}
This paper is a result of a project introduced by the Deanship of
Scientific Research (DSR) at King Abdulaziz University, Jeddah,
under grant no.(G-662-130-37). The authors, therefore, acknowledge
with thanks DSR for technical and financial support. The authors
thank the anonymous referee for his/her valuable comments.

\bibliographystyle{pasa-mnras}
\bibliography{PNe}

\newpage
\begin{appendix}
\section*{Appendix: Emission Line Fluxes and ionic abundances}

The observed and de-reddened line fluxes (relative to H$\beta$ =
100), log F(H$\beta$), and $ {c(H\beta)}$ of Hen\,2-141, NGC\,5307,
IC\,2553, and PB\,6 as well as their studied knots were given in
Tables \ref{TableA1} and \ref{TableA2}. The [O III] line at $\lambda
5007$ is saturated in NGC\,5307 and IC\,2553 nebulae. The values of
both lines were determined applying the theoretical line ratio [O
III] $\lambda 5007$ / [O III] $\lambda 4959$  of 2.89
\citep{Storey00}. Both of the [O III] lines at $\lambda 5007$ and
$\lambda 4959$ are saturated in Hen\,2-141, therefore we adopted
their values as the averages of available values in literatures.
Columns (1)and (2) give the laboratory wavelength and identification
of observed lines, while other columns show the observed and
de-reddened line fluxes of the PNe and their associated knots.

The total helium abundances for all objects were determined from
He$^{+}$/H and He$^{2+}$/H ions using atomic data from
\citet{Porter12} and \citet{Porter13}. The total carbon abundances
are determined from C$^{2+}$/H and C$^{3+}$/H ions in all objects,
except Hen\,2-141 which is determined from the C$^{2+}$/H ion only.
The ionic abundances of the four PNe and the knots were given in
Tables \ref{TableA3} and \ref{TableA4}.

\onecolumn
\begin{landscape}
\begin{table*}
\caption{Integrated line fluxes and de-reddened intensities,
relative to H$\beta = 100$, of Hen 2-141 and NGC 5307.}
\label{TableA1} \scalebox{0.67}{
\begin{tabular}{llllllllllllllll}

\hline
        &           & \multicolumn{2}{c}{Hen 2-141}  && \multicolumn{8}{c}{NGC 5307} \\
\cline{3-4} \cline{6-13}
\\

     &           & \multicolumn{2}{c}{Entire nebula}      &&    \multicolumn{2}{c}{Entire nebula}       &    \multicolumn{2}{c}{K$_{\textrm{NW}}$}                &      \multicolumn{2}{c}{K$_{\textrm{SW}}$}         &   \multicolumn{2}{c}{K$_{\textrm{SE}}$} \\

     $\lambda_{\rm Lab}$ \AA      &    ID              & \multicolumn{1}{c}{F($\lambda$)}  & \multicolumn{1}{c}{I($\lambda$)} &&  \multicolumn{1}{c}{F($\lambda$)} & \multicolumn{1}{c}{I($\lambda$)}  & \multicolumn{1}{c}{F($\lambda$)}  & \multicolumn{1}{c}{I($\lambda$)}& \multicolumn{1}{c}{F($\lambda$)}  & \multicolumn{1}{c}{I($\lambda$)}& \multicolumn{1}{c}{F($\lambda$)}  &\multicolumn{1}{c}{I($\lambda$)} \\
\hline
3703.86 &   H   I   &               &               &&  1.75    $\pm$   0.14    &   2.42    $\pm$   0.23    &   1.96    $\pm$   0.32    &   2.74    $\pm$   0.45    &   1.13    $\pm$   0.17    &   1.68    $\pm$   0.25    &               &               \\
3711.97 &   H   I   &   0.71    $\pm$   0.14    &   1.20    $\pm$   0.23    &&  1.18    $\pm$   0.11    &   1.63    $\pm$   0.16    &               &               &   0.71    $\pm$   0.13    &   1.06    $\pm$   0.20    &               &               \\
3726.03 &   [O  II] &   46.83   $\pm$   6.14    &   78.58   $\pm$   10.05   &&  10.03   $\pm$   0.25    &   13.85   $\pm$   0.82    &   81.16   $\pm$   3.82    &   113.04  $\pm$   6.38    &   15.52   $\pm$   0.41    &   22.97   $\pm$   0.93    &   29.13   $\pm$   4.17    &   42.91   $\pm$   6.24    \\
3728.82 &   [O  II] &   30.46   $\pm$   4.51    &   50.91   $\pm$   7.27    &&  5.14    $\pm$   0.22    &   7.08    $\pm$   0.47    &   42.88   $\pm$   3.74    &   59.35   $\pm$   5.30    &   8.79    $\pm$   0.43    &   12.98   $\pm$   0.73    &   15.66   $\pm$   2.24    &   23.02   $\pm$   3.36    \\
3734.37 &   H   I   &   1.00    $\pm$   0.21    &   1.70    $\pm$   0.35    &&  1.68    $\pm$   0.08    &   2.31    $\pm$   0.16    &   2.27    $\pm$   0.48    &   3.17    $\pm$   0.69    &   1.57    $\pm$   0.19    &   2.30    $\pm$   0.28    &   0.87    $\pm$   0.26    &   1.28    $\pm$   0.39    \\
3750.15 &   H   I   &   2.33    $\pm$   0.23    &   3.89    $\pm$   0.37    &&  2.19    $\pm$   0.09    &   3.00    $\pm$   0.20    &   3.34    $\pm$   0.22    &   4.62    $\pm$   0.32    &   1.35    $\pm$   0.12    &   1.99    $\pm$   0.18    &   0.73    $\pm$   0.22    &   1.07    $\pm$   0.32    \\
3770.63 &   H   I   &   2.70    $\pm$   0.24    &   4.48    $\pm$   0.39    &&  2.85    $\pm$   0.13    &   3.89    $\pm$   0.26    &   4.43    $\pm$   0.29    &   6.11    $\pm$   0.45    &   2.88    $\pm$   0.17    &   4.21    $\pm$   0.27    &   1.22    $\pm$   0.25    &   1.78    $\pm$   0.37    \\
3797.90 &   H   I   &   3.99    $\pm$   0.29    &   6.57    $\pm$   0.49    &&  3.85    $\pm$   0.09    &   5.24    $\pm$   0.30    &   6.07    $\pm$   0.42    &   8.29    $\pm$   0.60    &   3.87    $\pm$   0.19    &   5.62    $\pm$   0.32    &   1.62    $\pm$   0.29    &   2.34    $\pm$   0.40    \\
3834.89 &   He  II  &   1.31    $\pm$   0.32    &   2.13    $\pm$   0.52    &&  5.36    $\pm$   0.09    &   7.23    $\pm$   0.39    &               &               &               &               &   1.13    $\pm$   0.33    &   1.61    $\pm$   0.47    \\
3868.75 &   [Ne III]    &   74.05   $\pm$   5.23    &   118.41  $\pm$   8.45    &&  87.77   $\pm$   0.55    &   117.37  $\pm$   5.83    &   121.75  $\pm$   1.46    &   163.77  $\pm$   4.87    &   92.24   $\pm$   0.97    &   131.15  $\pm$   3.88    &   82.58   $\pm$   8.22    &   116.08  $\pm$   11.27   \\
3888.65 &   He  I   &   10.37   $\pm$   0.67    &   17.10   $\pm$   0.49    &&  14.65   $\pm$   0.12    &   19.50   $\pm$   0.95    &               &               &               &               &               &               \\
3967.46 &   [Ne III]    &   24.86   $\pm$   1.85    &   38.15   $\pm$   2.85    &&  24.76   $\pm$   0.43    &   32.29   $\pm$   1.56    &   36.74   $\pm$   0.83    &   48.19   $\pm$   1.63    &   26.21   $\pm$   0.52    &   36.15   $\pm$   1.16    &   26.21   $\pm$   3.13    &   35.66   $\pm$   4.12    \\
3970.07 &   H   I   &   9.99    $\pm$   1.54    &   15.36   $\pm$   2.41    &&  11.27   $\pm$   0.41    &   14.66   $\pm$   0.83    &   18.81   $\pm$   0.69    &   24.66   $\pm$   1.10    &   11.95   $\pm$   0.50    &   16.47   $\pm$   0.81    &   10.69   $\pm$   0.55    &   14.63   $\pm$   0.78    \\
4068.60 &   [S  II] &   1.70    $\pm$   0.10    &   4.12    $\pm$   0.06    &&              &               &               &               &               &               &               &               \\
4100.04 &   He  II  &   2.44    $\pm$   0.68    &   3.53    $\pm$   0.96    &&              &               &               &               &               &               &               &               \\
4101.74 &   H   I   &   19.11   $\pm$   0.68    &   27.71   $\pm$   1.01    &&  20.82   $\pm$   0.10    &   26.17   $\pm$   1.02    &   30.57   $\pm$   0.29    &   38.63   $\pm$   0.92    &   21.16   $\pm$   0.23    &   27.93   $\pm$   0.68    &   17.83   $\pm$   1.37    &   23.35   $\pm$   1.70    \\
4267.15 &   C   II  &   0.48    $\pm$   0.10    &   0.65    $\pm$   0.14    &&  0.13    $\pm$   0.02    &   0.15    $\pm$   0.03    &               &               &   0.20    $\pm$   0.04    &   0.25    $\pm$   0.05    &               &               \\
4340.47 &   H   I   &   36.93   $\pm$   0.30    &   47.91   $\pm$   0.60    &&  39.71   $\pm$   0.31    &   46.59   $\pm$   1.30    &   54.46   $\pm$   0.91    &   64.12   $\pm$   1.44    &   40.87   $\pm$   1.05    &   49.59   $\pm$   1.41    &   38.43   $\pm$   2.30    &   46.41   $\pm$   2.57    \\
4363.21 &   [O  III]    &   13.40   $\pm$   0.15    &   17.19   $\pm$   0.25    &&  13.74   $\pm$   0.23    &   16.01   $\pm$   0.49    &   12.75   $\pm$   0.29    &   14.91   $\pm$   0.40    &   13.35   $\pm$   0.35    &   16.07   $\pm$   0.48    &   12.15   $\pm$   0.71    &   14.56   $\pm$   0.84    \\
4471.50 &   He  I   &   2.11    $\pm$   0.12    &   2.56    $\pm$   0.15    &&  3.47    $\pm$   0.03    &   3.91    $\pm$   0.09    &   1.17    $\pm$   0.20    &   1.33    $\pm$   0.23    &   4.15    $\pm$   0.07    &   4.81    $\pm$   0.10    &   4.32    $\pm$   0.31    &   4.98    $\pm$   0.36    \\
4541.59 &   He  II  &   2.07    $\pm$   0.07    &   2.43    $\pm$   0.09    &&  0.73    $\pm$   0.03    &   0.80    $\pm$   0.03    &               &               &               &               &               &               \\
4634.14 &   N   III &   0.71    $\pm$   0.09    &   0.80    $\pm$   0.10    &&              &               &               &               &               &               &               &               \\
4640.64 &   N   III &   1.46    $\pm$   0.08    &   1.63    $\pm$   0.09    &&  0.96    $\pm$   0.02    &   1.03    $\pm$   0.02    &               &               &   0.55    $\pm$   0.02    &   0.60    $\pm$   0.02    &   0.55    $\pm$   0.03    &   0.60    $\pm$   0.03    \\
4641.81 &   O   II  &   0.28    $\pm$   0.08    &   0.51    $\pm$   0.00    &&  0.18    $\pm$   0.01    &   0.20    $\pm$   0.01    &               &               &               &               &               &               \\
4649.13 &   O   II  &   0.28    $\pm$   0.08    &   0.31    $\pm$   0.08    &&  0.14    $\pm$   0.02    &   0.15    $\pm$   0.02    &               &               &               &               &               &               \\
4650.25 &   C   III &               &               &&  0.15    $\pm$   0.02    &   0.17    $\pm$   0.02    &   0.13    $\pm$   0.03    &   0.14    $\pm$   0.04    &   0.15    $\pm$   0.03    &   0.16    $\pm$   0.03    &               &               \\
4685.68 &   He  II  &   63.61   $\pm$   0.52    &   69.50   $\pm$   0.62    &&  23.03   $\pm$   0.19    &   24.32   $\pm$   0.30    &   1.41    $\pm$   0.10    &   1.49    $\pm$   0.11    &   7.34    $\pm$   0.19    &   7.84    $\pm$   0.20    &   1.53    $\pm$   0.09    &   1.63    $\pm$   0.10    \\
4711.37 &   [Ar IV] &   7.28    $\pm$   0.17    &   7.85    $\pm$   0.19    &&  4.79    $\pm$   0.10    &   5.02    $\pm$   0.12    &   2.12    $\pm$   0.11    &   2.22    $\pm$   0.11    &   3.39    $\pm$   0.09    &   3.59    $\pm$   0.09    &   2.95    $\pm$   0.17    &   3.11    $\pm$   0.18    \\
4724.89 &   [Ne IV] &   1.73    $\pm$   0.10    &   1.86    $\pm$   0.11    &&              &               &               &               &               &               &               &               \\
4740.17 &   [Ar IV] &   6.26    $\pm$   0.10    &   6.66    $\pm$   0.11    &&  4.14    $\pm$   0.07    &   4.30    $\pm$   0.07    &   1.84    $\pm$   0.10    &   1.91    $\pm$   0.10    &   3.02    $\pm$   0.09    &   3.16    $\pm$   0.09    &   2.60    $\pm$   0.14    &   2.72    $\pm$   0.15    \\
4861.33 &   H   I   &   100.00  $\pm$   1.64    &   100.00  $\pm$   1.64    &&  100.00  $\pm$   2.94    &   100.00  $\pm$   2.94    &   100.00  $\pm$   4.44    &   100.00  $\pm$   4.46    &   100.00  $\pm$   6.19    &   100.00  $\pm$   6.13    &   100.00  $\pm$   2.37    &   100.00  $\pm$   2.36    \\
4921.93 &   He  I   &   0.64    $\pm$   0.14    &   0.62    $\pm$   0.14    &&  1.13    $\pm$   0.04    &   1.11    $\pm$   0.04    &   1.81    $\pm$   0.08    &   1.78    $\pm$   0.08    &   1.23    $\pm$   0.10    &   1.20    $\pm$   0.09    &   1.24    $\pm$   0.05    &   1.22    $\pm$   0.05    \\
4958.91 &   [O  III]    &   519.98  $\pm$   1.65    &   495.00  $\pm$   1.85    &&  459.97  $\pm$   42.79   &   444.00  $\pm$   40.39   &   426.78  $\pm$   18.09   &   414.00  $\pm$   17.70   &   469.92  $\pm$   35.11   &   453.00  $\pm$   33.89   &   452.25  $\pm$   36.98   &   436.00  $\pm$   35.64   \\
5006.84 &   [O  III]    &   1606.35 $\pm$   23.19   &   1493.00 $\pm$   22.00   &&  1402.29 $\pm$   134.23  &   1336.00 $\pm$   128.00  &   1268.20 $\pm$   46.90   &   1211.00 $\pm$   45.00   &   1503.03 $\pm$   22.86   &   1464.00 $\pm$   22.00   &   1249.39 $\pm$   52.73   &   1184.00 $\pm$   50.00   \\
5015.68 &   He  I   &   3.37    $\pm$   0.14    &   3.12    $\pm$   0.13    &&  1.83    $\pm$   0.33    &   1.75    $\pm$   0.32    &   1.23    $\pm$   0.32    &   1.17    $\pm$   0.30    &               &               &   2.18    $\pm$   0.53    &   2.05    $\pm$   0.50    \\
5199.84 &   [N  I]  &   2.31    $\pm$   0.33    &   3.20    $\pm$   0.02    &&  0.10    $\pm$   0.03    &   0.09    $\pm$   0.03    &               &               &   0.15    $\pm$   0.01    &   0.13    $\pm$   0.11    &   0.24    $\pm$   0.02    &   0.22    $\pm$   0.02    \\
5200.26 &   [N  I]  &               &               &&  0.08    $\pm$   0.03    &   0.07    $\pm$   0.03    &   0.50    $\pm$   0.14    &   0.45    $\pm$   0.12    &   0.08    $\pm$   0.01    &   0.07    $\pm$   0.07    &   0.30    $\pm$   0.02    &   0.26    $\pm$   0.02    \\
5411.52 &   He  II  &   7.10    $\pm$   0.39    &   5.40    $\pm$   0.30    &&  2.28    $\pm$   0.21    &   1.92    $\pm$   0.18    &               &               &   0.82    $\pm$   0.07    &   0.66    $\pm$   0.05    &   0.19    $\pm$   0.04    &   0.16    $\pm$   0.03    \\
5517.66 &   [Cl III]    &   1.33    $\pm$   0.29    &   0.97    $\pm$   0.22    &&  0.87    $\pm$   0.07    &   0.71    $\pm$   0.06    &   1.55    $\pm$   0.51    &   1.28    $\pm$   0.41    &   0.58    $\pm$   0.05    &   0.46    $\pm$   0.04    &   0.89    $\pm$   0.24    &   0.71    $\pm$   0.19    \\
5537.60 &   [Cl III]    &   1.18    $\pm$   0.23    &   0.86    $\pm$   0.16    &&  0.80    $\pm$   0.06    &   0.65    $\pm$   0.05    &   1.42    $\pm$   0.54    &   1.15    $\pm$   0.43    &   0.60    $\pm$   0.07    &   0.47    $\pm$   0.06    &   0.86    $\pm$   0.20    &   0.68    $\pm$   0.16    \\
5754.60 &   [N  II] &   3.28    $\pm$   0.32    &   2.20    $\pm$   0.21    &&  0.38    $\pm$   0.05    &   0.30    $\pm$   0.04    &   3.41    $\pm$   0.14    &   2.66    $\pm$   0.12    &   1.05    $\pm$   0.05    &   0.78    $\pm$   0.04    &   1.51    $\pm$   0.05    &   1.13    $\pm$   0.04    \\
5801.51 &   C   IV  &   0.94    $\pm$   0.19    &   0.62    $\pm$   0.13    &&  0.14    $\pm$   0.03    &   0.11    $\pm$   0.02    &               &               &               &               &               &               \\
5812.14 &   C   IV  &               &               &&  0.08    $\pm$   0.02    &   0.07    $\pm$   0.01    &               &               &   0.10    $\pm$   0.03    &   0.08    $\pm$   0.02    &               &               \\
5875.66 &   He  I   &   12.93   $\pm$   0.50    &   8.33    $\pm$   0.35    &&  16.07   $\pm$   1.85    &   12.20   $\pm$   1.45    &   23.90   $\pm$   0.19    &   18.14   $\pm$   0.49    &   26.78   $\pm$   0.14    &   19.31   $\pm$   0.51    &   19.39   $\pm$   0.06    &   14.06   $\pm$   0.20    \\
6300.34 &   [O  I]  &   13.99   $\pm$   1.11    &   7.83    $\pm$   0.62    &&  2.11    $\pm$   0.22    &   1.47    $\pm$   0.17    &   19.91   $\pm$   1.00    &   13.84   $\pm$   0.81    &   6.90    $\pm$   0.51    &   4.48    $\pm$   0.35    &   8.44    $\pm$   0.26    &   5.54    $\pm$   0.20    \\
6312.10 &   [S  III]    &   4.41    $\pm$   0.20    &   2.46    $\pm$   0.12    &&  1.73    $\pm$   0.13    &   1.20    $\pm$   0.11    &   6.32    $\pm$   0.44    &   4.38    $\pm$   0.32    &   2.82    $\pm$   0.14    &   1.83    $\pm$   0.11    &   3.14    $\pm$   0.24    &   2.06    $\pm$   0.16    \\
6363.78 &   [O  I]  &   4.74    $\pm$   0.40    &   2.60    $\pm$   0.22    &&  0.79    $\pm$   0.09    &   0.54    $\pm$   0.07    &   7.28    $\pm$   0.25    &   5.01    $\pm$   0.24    &   2.32    $\pm$   0.17    &   1.48    $\pm$   0.12    &   2.48    $\pm$   0.13    &   1.60    $\pm$   0.09    \\
6548.10 &   [N  II] &   72.19   $\pm$   3.73    &   37.52   $\pm$   2.12    &&  5.02    $\pm$   0.44    &   3.36    $\pm$   0.35    &   46.61   $\pm$   3.33    &   30.87   $\pm$   2.42    &   12.60   $\pm$   0.82    &   7.73    $\pm$   0.56    &   21.29   $\pm$   2.25    &   13.14   $\pm$   1.37    \\
6562.77 &   H   I   &   548.05  $\pm$   13.47   &   284.00  $\pm$   3.40    &&  421.82  $\pm$   31.84   &   281.00  $\pm$   5.60    &   458.98  $\pm$   3.46    &   304.00  $\pm$   11.60   &   467.07  $\pm$   1.50    &   287.00  $\pm$   11.10   &   450.30  $\pm$   2.20    &   279.00  $\pm$   5.50    \\
6583.50 &   [N  II] &   216.38  $\pm$   21.24   &   111.10  $\pm$   10.93   &&  14.67   $\pm$   1.59    &   9.72    $\pm$   1.19    &   144.46  $\pm$   13.78   &   94.84   $\pm$   9.45    &   40.94   $\pm$   1.80    &   25.02   $\pm$   1.47    &   66.38   $\pm$   5.13    &   40.74   $\pm$   3.14    \\
6678.16 &   He  I   &   3.98    $\pm$   0.13    &   1.99    $\pm$   0.08    &&  4.94    $\pm$   0.32    &   3.23    $\pm$   0.30    &   10.02   $\pm$   0.12    &   6.50    $\pm$   0.27    &   7.86    $\pm$   0.51    &   4.70    $\pm$   0.35    &   6.18    $\pm$   0.27    &   3.74    $\pm$   0.18    \\
6716.44 &   [S  II] &   11.66   $\pm$   0.25    &   5.78    $\pm$   0.20    &&  2.07    $\pm$   0.14    &   1.35    $\pm$   0.13    &   20.86   $\pm$   0.90    &   13.43   $\pm$   0.77    &   4.71    $\pm$   0.20    &   2.80    $\pm$   0.16    &   8.16    $\pm$   0.32    &   4.89    $\pm$   0.22    \\
6730.82 &   [S  II] &   15.35   $\pm$   0.36    &   7.58    $\pm$   0.27    &&  3.49    $\pm$   0.25    &   2.26    $\pm$   0.22    &   28.76   $\pm$   1.26    &   18.47   $\pm$   1.07    &   7.63    $\pm$   0.47    &   4.50    $\pm$   0.32    &   13.11   $\pm$   0.60    &   7.83    $\pm$   0.39    \\
6890.88 &   He  I   &   2.06    $\pm$   0.19    &   0.97    $\pm$   0.09    &&  0.46    $\pm$   0.04    &   0.29    $\pm$   0.03    &               &               &       &               &               &                       \\
7005.67 &   [Ar V]  &   2.94    $\pm$   0.24    &   1.35    $\pm$   0.11    &&  0.37    $\pm$   0.03    &   0.23    $\pm$   0.03    &               &               &       &               &               &                       \\

\hline
${c(H\beta)}$  &   &  \multicolumn{2}{c}{0.89} && \multicolumn{2}{c}{0.54} & \multicolumn{2}{c}{0.56}  & \multicolumn{2}{c}{0.66} & \multicolumn{2}{c}{0.65}  \\
F$(H\beta)$   &   &  \multicolumn{2}{c}{-11.78} && \multicolumn{2}{c}{-11.29} & \multicolumn{2}{c}{-13.0}  & \multicolumn{2}{c}{-12.55} & \multicolumn{2}{c}{-12.78}  \\
\hline
\end{tabular}}
\end{table*}
\end{landscape}

\begin{landscape}
\begin{table*}
\centering \caption{Integrated line fluxes and de-reddened
intensities, relative to H$\beta = 100$, of IC 2553 and PB6.}
\label{TableA2} \scalebox{0.62}{
\begin{tabular}{llllllllllllllllll}
\hline
        &           & \multicolumn{6}{c}{IC 2553}  && \multicolumn{4}{c}{PB6} \\
\cline{3-8} \cline{10-13}
       &           & \multicolumn{2}{c}{Entire nebula}  &  \multicolumn{2}{c}{K$_{\textrm{NE}}$} &    \multicolumn{2}{c}{K$_{\textrm{SW}}$}  &&      \multicolumn{2}{c}{Entire nebula}   &   \multicolumn{2}{c}{K$_{\textrm{SE}}$} \\

      $\lambda_{\rm Lab}$ \AA      &    ID            & \multicolumn{1}{c}{F($\lambda$)}  & \multicolumn{1}{c}{I($\lambda$)} &  \multicolumn{1}{c}{F($\lambda$)}  & \multicolumn{1}{c}{I($\lambda$)} & \multicolumn{1}{c}{F($\lambda$)}  & \multicolumn{1}{c}{I($\lambda$)}&& \multicolumn{1}{c}{F($\lambda$)}  & \multicolumn{1}{c}{I($\lambda$)}& \multicolumn{1}{c}{F($\lambda$)} &\multicolumn{1}{c}{I($\lambda$)} \\
\hline
3703.86 &   H   I   &   1.68    $\pm$   0.29    &   2.16    $\pm$   0.38    &               &               &               &               &&  1.08    $\pm$   0.22    &   1.55    $\pm$   0.32    &               &               \\
3726.03 &   [O  II] &   14.50   $\pm$   1.36    &   18.47   $\pm$   1.88    &   43.41   $\pm$   3.91    &   58.14   $\pm$   5.19    &   54.86   $\pm$   4.94    &   70.65   $\pm$   6.29    &&  30.76   $\pm$   1.18    &   44.13   $\pm$   1.86    &   56.38   $\pm$   5.65    &   75.20   $\pm$   7.54    \\
3728.82 &   [O  II] &   6.96    $\pm$   1.24    &   8.88    $\pm$   1.59    &   22.20   $\pm$   1.89    &   29.68   $\pm$   2.52    &   29.94   $\pm$   2.69    &   38.80   $\pm$   3.65    &&  19.36   $\pm$   1.18    &   27.78   $\pm$   1.76    &   36.57   $\pm$   3.29    &   49.07   $\pm$   4.52    \\
3734.37 &   H   I   &   2.11    $\pm$   0.22    &   2.70    $\pm$   0.29    &               &               &               &   3.16    $\pm$   0.28    &&  1.72    $\pm$   0.19    &   2.45    $\pm$   0.27    &               &               \\
3750.15 &   H   I   &   2.58    $\pm$   0.22    &   3.29    $\pm$   0.30    &   3.99    $\pm$   0.36    &   5.32    $\pm$   0.48    &   3.61    $\pm$   0.33    &   4.63    $\pm$   0.42    &&  1.45    $\pm$   0.21    &   2.07    $\pm$   0.31    &               &               \\
3770.63 &   H   I   &   3.24    $\pm$   0.24    &   4.11    $\pm$   0.35    &   3.84    $\pm$   0.35    &   5.10    $\pm$   0.46    &   3.80    $\pm$   0.34    &   4.86    $\pm$   0.43    &&  2.60    $\pm$   0.22    &   3.69    $\pm$   0.31    &   3.58    $\pm$   0.32    &   4.74    $\pm$   0.42    \\
3797.90 &   H   I   &   4.44    $\pm$   0.28    &   5.61    $\pm$   0.42    &   5.62    $\pm$   0.51    &   7.41    $\pm$   0.66    &   5.09    $\pm$   0.46    &   6.47    $\pm$   0.58    &&  5.20    $\pm$   0.63    &   7.34    $\pm$   0.89    &   3.83    $\pm$   0.34    &   5.04    $\pm$   0.45    \\
3834.89 &   He  II  &   6.33    $\pm$   0.24    &   7.95    $\pm$   0.46    &   7.44    $\pm$   0.67    &   9.74    $\pm$   0.87    &               &               &&  6.45    $\pm$   0.68    &   10.54   $\pm$   0.16    &   6.00    $\pm$   0.54    &   7.84    $\pm$   0.70    \\
3868.75 &   [Ne III]    &   86.05   $\pm$   3.47    &   107.35  $\pm$   6.23    &   114.33  $\pm$   10.98   &   148.26  $\pm$   13.90   &   125.68  $\pm$   6.28    &   158.26  $\pm$   8.59    &&  85.91   $\pm$   1.11    &   118.84  $\pm$   2.38    &   101.99  $\pm$   5.10    &   132.94  $\pm$   7.14    \\
3888.65 &   He  I   &   16.65   $\pm$   0.49    &   20.72   $\pm$   1.06    &   20.06   $\pm$   1.81    &   25.93   $\pm$   2.31    &   20.72   $\pm$   1.86    &   25.92   $\pm$   2.33    &&  13.01   $\pm$   0.41    &   20.97   $\pm$   0.31    &   15.54   $\pm$   1.40    &   20.09   $\pm$   1.78    \\
3967.46 &   [Ne III]    &   32.18   $\pm$   0.79    &   39.45   $\pm$   1.87    &   37.77   $\pm$   2.27    &   47.98   $\pm$   2.93    &   39.38   $\pm$   3.54    &   48.44   $\pm$   4.34    &&  22.15   $\pm$   0.33    &   29.80   $\pm$   0.60    &   23.59   $\pm$   2.12    &   29.94   $\pm$   2.68    \\
3970.07 &   H   I   &   12.41   $\pm$   0.68    &   15.17   $\pm$   1.01    &   13.97   $\pm$   1.26    &   17.69   $\pm$   1.56    &   15.28   $\pm$   1.38    &   18.82   $\pm$   1.67    &&  12.06   $\pm$   0.30    &   16.21   $\pm$   0.46    &   12.35   $\pm$   1.11    &   15.73   $\pm$   1.45    \\
4026.08 &   N   II  &   2.01    $\pm$   0.04    &   2.44    $\pm$   0.10    &   2.35    $\pm$   0.21    &   2.94    $\pm$   0.26    &   2.58    $\pm$   0.23    &   3.14    $\pm$   0.28    &&  1.96    $\pm$   0.23    &   3.03    $\pm$   0.04    &               &               \\
4068.60 &   [S  II] &   1.60    $\pm$   0.04    &   1.93    $\pm$   0.09    &   3.44    $\pm$   0.31    &   4.28    $\pm$   0.39    &   4.53    $\pm$   0.41    &   5.48    $\pm$   0.51    &&              &               &   2.99    $\pm$   0.27    &   3.69    $\pm$   0.33    \\
4076.35 &   [S  II] &   0.71    $\pm$   0.03    &   0.85    $\pm$   0.05    &   1.25        0.11    &               &   1.75    $\pm$   0.16    &   2.10    $\pm$   0.19    &&              &               &               &               \\
4100.04 &   He  II  &   0.29    $\pm$   0.08    &   0.34    $\pm$   0.10    &               &               &               &               &&  1.67    $\pm$   0.20    &   2.15    $\pm$   0.26    &               &               \\
4101.74 &   H   I   &   22.09   $\pm$   0.30    &   26.34   $\pm$   0.98    &   25.07   $\pm$   1.50    &   30.88   $\pm$   1.86    &   25.73   $\pm$   1.29    &   30.86   $\pm$   1.55    &&  21.10   $\pm$   0.23    &   27.25   $\pm$   0.43    &   22.48   $\pm$   1.12    &   27.69   $\pm$   1.38    \\
4199.83 &   He  II  &   0.47    $\pm$   0.06    &   0.55    $\pm$   0.08    &               &               &               &               &&  2.29    $\pm$   0.08    &   3.36    $\pm$   0.04    &   1.52    $\pm$   0.14    &   1.82    $\pm$   0.16    \\
4267.15 &   C   II  &   0.54    $\pm$   0.01    &   0.63    $\pm$   0.02    &   0.52    $\pm$   0.10    &   0.62    $\pm$   0.12    &   0.48    $\pm$   0.10    &   0.56    $\pm$   0.11    &&  0.36    $\pm$   0.08    &   0.44    $\pm$   0.09    &               &               \\
4340.47 &   H   I   &   40.87   $\pm$   0.40    &   46.21   $\pm$   1.20    &   37.68   $\pm$   2.26    &   43.57   $\pm$   2.45    &   46.53   $\pm$   2.33    &   52.82   $\pm$   2.55    &&  38.98   $\pm$   0.44    &   46.60   $\pm$   0.62    &   40.46   $\pm$   2.02    &   46.85   $\pm$   2.18    \\
4363.21 &   [O  III]    &   9.30    $\pm$   0.11    &   10.46   $\pm$   0.27    &   8.61    $\pm$   0.77    &   9.85    $\pm$   0.85    &   9.28    $\pm$   0.84    &   10.42   $\pm$   0.92    &&  16.62   $\pm$   0.32    &   19.72   $\pm$   0.41    &   15.73   $\pm$   1.42    &   18.01   $\pm$   1.60    \\
4471.50 &   He  I   &   4.22    $\pm$   0.04    &   4.63    $\pm$   0.09    &   4.96    $\pm$   0.45    &   5.52    $\pm$   0.48    &   5.35    $\pm$   0.48    &   5.89    $\pm$   0.53    &&  2.03    $\pm$   0.09    &   2.33    $\pm$   0.11    &   3.68    $\pm$   0.33    &   4.09    $\pm$   0.36    \\
4541.59 &   He  II  &   0.78    $\pm$   0.03    &   0.85    $\pm$   0.03    &   0.65        0.06    &               &               &               &&  4.38    $\pm$   0.09    &   4.89    $\pm$   0.11    &   3.77    $\pm$   0.34    &   4.12    $\pm$   0.37    \\
4634.14 &   N   III &   1.77    $\pm$   0.08    &   1.87    $\pm$   0.09    &   1.10    $\pm$   0.10    &   1.17    $\pm$   0.10    &   0.61    $\pm$   0.05    &   0.64    $\pm$   0.06    &&  1.04    $\pm$   0.14    &   1.12    $\pm$   0.14    &               &               \\
4640.64 &   N   III &   3.64    $\pm$   0.05    &   3.83    $\pm$   0.06    &   2.62    $\pm$   0.24    &   2.78    $\pm$   0.24    &   2.05    $\pm$   0.18    &   2.16    $\pm$   0.20    &&  1.42    $\pm$   0.08    &   1.54    $\pm$   0.09    &               &               \\
4647.42 &   C   III &               &               &               &               &               &               &&  0.45    $\pm$   0.07    &   0.49    $\pm$   0.08    &               &               \\
4650.25 &   C   III &   0.92    $\pm$   0.06    &   0.96    $\pm$   0.06    &   0.83    $\pm$   0.07    &   0.88    $\pm$   0.08    &               &               &&              &               &               &               \\
4658.64 &   C   IV  &   0.20    $\pm$   0.03    &   0.21    $\pm$   0.03    &               &               &               &               &&              &               &               &               \\
4685.68 &   He  II  &   24.14   $\pm$   0.38    &   25.17   $\pm$   0.45    &   14.51   $\pm$   1.31    &   15.23   $\pm$   1.37    &   9.49    $\pm$   0.28    &   9.91    $\pm$   0.30    &&  129.07  $\pm$   0.77    &   137.17  $\pm$   0.91    &   96.84   $\pm$   2.91    &   101.76  $\pm$   3.06    \\
4711.37 &   [Ar IV] &   5.39    $\pm$   0.07    &   5.59    $\pm$   0.08    &   2.82    $\pm$   0.25    &   2.95    $\pm$   0.26    &   2.55    $\pm$   0.23    &   2.66    $\pm$   0.24    &&  9.76    $\pm$   0.22    &   10.28   $\pm$   0.23    &   7.16    $\pm$   0.72    &   7.44    $\pm$   0.72    \\
4715.21 &   [Ne IV] &               &               &               &               &               &               &&  1.62    $\pm$   0.20    &   1.70    $\pm$   0.20    &               &               \\
4724.15 &   [Ne IV] &   0.28        0.04    &               &               &               &               &               &&  1.65    $\pm$   0.08    &   1.74    $\pm$   0.09    &               &               \\
4725.62 &   [Ne IV] &               &               &               &               &               &               &&  1.34    $\pm$   0.07    &   1.40    $\pm$   0.07    &               &               \\
4740.17 &   [Ar IV] &   6.03    $\pm$   0.11    &   6.21    $\pm$   0.12    &   2.91    $\pm$   0.26    &   3.02    $\pm$   0.27    &   2.63    $\pm$   0.28    &   2.71    $\pm$   0.29    &&  8.85    $\pm$   0.18    &   9.23    $\pm$   0.18    &   6.31    $\pm$   0.57    &   6.50    $\pm$   0.56    \\
4859.32 &   He  II  &               &               &               &               &               &               &&  6.76    $\pm$   0.85    &   6.78    $\pm$   0.85    &               &               \\
4861.33 &   H   I   &   100.00  $\pm$   3.82    &   100.00  $\pm$   3.86    &   100.00  $\pm$   3.00    &   100.00  $\pm$   2.98    &   100.00  $\pm$   3.00    &   100.00  $\pm$   2.96    &&  100.00  $\pm$   0.86    &   100.00  $\pm$   0.85    &   100.00  $\pm$   3.00    &   100.00  $\pm$   3.04    \\
4958.91 &   [O  III]    &   444.84  $\pm$   16.22   &   434.39  $\pm$   15.79   &   419.56  $\pm$   37.76   &   406.66  $\pm$   35.66   &   412.84  $\pm$   12.39   &   403.07  $\pm$   12.16   &&  365.24  $\pm$   2.12    &   353.13  $\pm$   2.13    &   379.38  $\pm$   11.38   &   369.00  $\pm$   11.20   \\
5006.84 &   [O  III]    &   1289.98 $\pm$   46.98   &   1245.55 $\pm$   45.94   &   1220.10 $\pm$   36.60   &   1170.78 $\pm$   35.42   &   1240.62 $\pm$   37.22   &   1197.60 $\pm$   36.13   &&  1098.89 $\pm$   4.35    &   1045.01 $\pm$   4.78    &   1199.83 $\pm$   35.99   &   1151.54 $\pm$   34.37   \\
5191.82 &   [Ar III]    &   0.14    $\pm$   0.01    &   0.13    $\pm$   0.01    &   0.16    $\pm$   0.01    &   0.15    $\pm$   0.01    &               &               &&              &               &               &               \\
5197.81 &   [N  I]  &   0.14    $\pm$   0.01    &   0.13    $\pm$   0.01    &   0.28    $\pm$   0.03    &   0.25    $\pm$   0.02    &   0.72    $\pm$   0.07    &   0.66    $\pm$   0.06    &&  1.20    $\pm$   0.34    &   1.07    $\pm$   0.31    &   3.07        0.28    &               \\
5200.26 &   [N  I]  &   0.10    $\pm$   0.03    &   0.09    $\pm$   0.02    &   0.22    $\pm$   0.02    &   0.20    $\pm$   0.02    &   0.62    $\pm$   0.06    &   0.57    $\pm$   0.05    &&  1.03    $\pm$   0.34    &   0.92    $\pm$   0.31    &   1.98        0.18    &               \\
5411.52 &   He  II  &   2.16    $\pm$   0.13    &   1.90    $\pm$   0.12    &   1.19    $\pm$   0.11    &   1.02    $\pm$   0.09    &   0.59    $\pm$   0.05    &   0.51    $\pm$   0.05    &&  14.02   $\pm$   0.95    &   11.63   $\pm$   0.79    &   9.47    $\pm$   0.85    &   8.10    $\pm$   0.71    \\
5517.66 &   [Cl III]    &   0.77    $\pm$   0.03    &   0.66    $\pm$   0.03    &   0.78    $\pm$   0.07    &   0.65    $\pm$   0.06    &   0.89    $\pm$   0.08    &   0.76    $\pm$   0.07    &&  1.05    $\pm$   0.22    &   0.85    $\pm$   0.18    &   0.74    $\pm$   0.06    &   0.62    $\pm$   0.05    \\
5537.60 &   [Cl III]    &   0.99    $\pm$   0.09    &   0.84    $\pm$   0.08    &   0.91    $\pm$   0.08    &   0.76    $\pm$   0.07    &   0.93    $\pm$   0.07    &   0.79    $\pm$   0.06    &&  1.04    $\pm$   0.29    &   0.83    $\pm$   0.23    &   0.73    $\pm$   0.10    &   0.61    $\pm$   0.08    \\
5592.37 &   O   III &   0.09    $\pm$   0.01    &   0.08    $\pm$   0.01    &               &               &               &               &&              &               &               &               \\
5754.60 &   [N  II] &   0.90    $\pm$   0.09    &   0.74    $\pm$   0.08    &   2.43    $\pm$   0.15    &   1.94    $\pm$   0.12    &   2.36    $\pm$   0.21    &   1.93    $\pm$   0.17    &&  6.63    $\pm$   0.42    &   5.03    $\pm$   0.32    &   9.27    $\pm$   0.75    &   7.39    $\pm$   0.59    \\
5801.51 &   C   IV  &   0.08    $\pm$   0.01    &   0.07    $\pm$   0.01    &               &               &               &               &&  0.76    $\pm$   0.13    &   0.57    $\pm$   0.10    &               &               \\
5812.14 &   C   IV  &   0.05    $\pm$   0.01    &   0.04    $\pm$   0.01    &               &               &               &               &&  0.53    $\pm$   0.14    &   0.40    $\pm$   0.10    &               &               \\
5875.66 &   He  I   &   16.88   $\pm$   1.51    &   13.66   $\pm$   1.29    &   12.12   $\pm$   1.09    &   9.44    $\pm$   0.85    &   17.53   $\pm$   1.58    &   14.08   $\pm$   1.25    &&  10.73   $\pm$   0.57    &   7.92    $\pm$   0.44    &   12.39   $\pm$   1.12    &   9.64    $\pm$   0.86    \\
6300.34 &   [O  I]  &   1.86    $\pm$   0.10    &   1.41    $\pm$   0.11    &   4.70    $\pm$   0.42    &   3.39    $\pm$   0.30    &   5.81    $\pm$   0.52    &   4.37    $\pm$   0.40    &&  6.63    $\pm$   0.18    &   4.46    $\pm$   0.15    &   9.46    $\pm$   0.85    &   6.82    $\pm$   0.62    \\
6312.10 &   [S  III]    &   3.05    $\pm$   0.21    &   2.31    $\pm$   0.20    &   2.49    $\pm$   0.22    &   1.79    $\pm$   0.16    &   3.86    $\pm$   0.35    &   2.90    $\pm$   0.28    &&  4.78    $\pm$   0.14    &   3.21    $\pm$   0.11    &   3.39    $\pm$   0.31    &   2.44    $\pm$   0.22    \\
6363.78 &   [O  I]  &   0.61    $\pm$   0.04    &   0.46    $\pm$   0.04    &   1.61    $\pm$   0.05    &   1.16    $\pm$   0.05    &   2.04    $\pm$   0.18    &   1.52    $\pm$   0.14    &&  2.29    $\pm$   0.07    &   1.52    $\pm$   0.05    &   3.47    $\pm$   0.31    &   2.47    $\pm$   0.22    \\
6548.10 &   [N  II] &   17.18   $\pm$   1.14    &   12.61   $\pm$   1.09    &   51.28   $\pm$   0.95    &   35.61   $\pm$   1.16    &   44.33   $\pm$   1.33    &   32.23   $\pm$   1.42    &&  119.83  $\pm$   4.09    &   76.58   $\pm$   3.02    &   184.52  $\pm$   5.54    &   128.17  $\pm$   5.31    \\
6560.10 &   He  II  &               &               &               &               &               &               &&  28.54   $\pm$   7.49    &   18.11   $\pm$   4.71    &   13.30   $\pm$   1.20    &   9.16    $\pm$   0.82    \\
6562.77 &   H   I   &   386.61  $\pm$   25.75   &   283.14  $\pm$   6.50    &   410.39  $\pm$   7.15    &   284.34  $\pm$   6.38    &   409.33  $\pm$   11.24   &   296.69  $\pm$   6.61    &&  438.20  $\pm$   9.86    &   279.12  $\pm$   1.75    &   406.06  $\pm$   8.98    &   281.26  $\pm$   6.03    \\
6583.50 &   [N  II] &   46.16   $\pm$   3.24    &   33.69   $\pm$   3.03    &   143.64  $\pm$   4.86    &   99.18   $\pm$   4.28    &   142.17  $\pm$   4.27    &   102.82  $\pm$   4.54    &&  340.81  $\pm$   8.30    &   216.47  $\pm$   7.01    &   534.80  $\pm$   1.00    &   369.29  $\pm$   10.71   \\
6678.16 &   He  I   &   5.08    $\pm$   0.30    &   3.66    $\pm$   0.31    &   3.35    $\pm$   0.30    &   2.27    $\pm$   0.21    &   6.71    $\pm$   0.34    &   4.78    $\pm$   0.28    &&  3.23    $\pm$   0.13    &   2.01    $\pm$   0.09    &   4.36    $\pm$   0.39    &   2.95    $\pm$   0.27    \\
6716.44 &   [S  II] &   4.28    $\pm$   0.25    &   3.08    $\pm$   0.26    &   9.07    $\pm$   0.54    &   6.12    $\pm$   0.40    &   13.88   $\pm$   0.69    &   9.83    $\pm$   0.58    &&  10.10   $\pm$   0.26    &   6.24    $\pm$   0.21    &   13.97   $\pm$   0.70    &   9.43    $\pm$   0.54    \\
6730.82 &   [S  II] &   7.08    $\pm$   0.38    &   5.07    $\pm$   0.42    &   14.89   $\pm$   0.89    &   10.03   $\pm$   0.66    &   19.50   $\pm$   1.76    &   13.76   $\pm$   1.27    &&  14.22   $\pm$   0.38    &   8.77    $\pm$   0.31    &   18.79   $\pm$   0.97    &   12.67   $\pm$   0.76    \\
6890.88 &   He  I   &   0.33    $\pm$   0.03    &   0.23    $\pm$   0.03    &               &               &               &               &&  2.18    $\pm$   0.09    &   1.30    $\pm$   0.06    &               &               \\
7005.67 &   [Ar V]  &   0.83    $\pm$   0.09    &   0.57    $\pm$   0.07    &               &               &               &               &&  7.00    $\pm$   0.18    &   4.10    $\pm$   0.14    &   2.08    $\pm$   0.19    &   1.34    $\pm$   0.12    \\

\hline
${c(H\beta)}$  &   &  \multicolumn{2}{c}{0.42} & \multicolumn{2}{c}{0.50} & \multicolumn{2}{c}{0.44}  && \multicolumn{2}{c}{0.61} & \multicolumn{2}{c}{0.50}  \\
F$(H\beta)$   &   &  \multicolumn{2}{c}{-10.88} & \multicolumn{2}{c}{-13.4} & \multicolumn{2}{c}{-12.6}  && \multicolumn{2}{c}{-11.93} & \multicolumn{2}{c}{-13.3}  \\
\hline
\end{tabular}}
 \end{table*}
\end{landscape}

\begin{table*}
\centering \caption{Ionic and total abundances of Hen 2-141 and NGC
5307.} \label{TableA3} \scalebox{0.65}{
\begin{tabular}{lccccccc}

\hline
    &  \multicolumn{1}{c}{Hen 2-141}  && \multicolumn{4}{c}{NGC 5307}\\    \cline{2-2} \cline{4-7}
\\
                        &       Entire PN                           &&     Entire PN                            &       K$_{\textrm{NW}}$                                  &       K$_{\textrm{SW}}$                                  &       K$_{\textrm{SE}}$      \\
\hline
\\
He$^{+}$/H          &   5.35E-02    (   2.6E-3  )   (   -2.6E-3 )   &&  8.04E-02    (   7.2E-3  )   (   -6.6E-3 )   &   1.09E-01    (   6.0E-3  )   (   -5.5E-3 )   &   1.17E-01    (   4.3E-3  )   (   -4.3E-3 )   &   9.28E-02    (   5.0E-3  )   (   -4.8E-3 )   \\
He$^{2+}$/H         &   5.86E-02    (   5.8E-4  )   (   -5.8E-4 )   &&  2.06E-02    (   3.0E-4  )   (   -3.0E-4 )   &   1.26E-03    (   9.1E-5  )   (   -9.1E-5 )   &   6.70E-03    (   1.7E-4  )   (   -1.7E-4 )   &   1.42E-03    (   8.6E-5  )   (   -8.1E-5 )   \\
He/H                &   1.12E-01    (   2.5E-3  )   (   -2.5E-3 )   &&  1.01E-01    (   7.0E-3  )   (   -6.6E-3 )   &   1.10E-01    (   6.0E-3  )   (   -5.5E-3 )   &   1.24E-01    (   4.3E-3  )   (   -4.3E-3 )   &   9.43E-02    (   5.0E-3  )   (   -4.8E-3 )   \\
\\
C$^{2+}$/H          &   6.52E-04    (   1.4E-4  )   (   -1.4E-4 )   &&  1.52E-04    (   2.7E-5  )   (   -2.6E-5 )   &                               &   2.52E-04    (   5.3E-5  )   (   -5.2E-5 )   &                               \\
C$^{3+}$/H          &                               &&  1.43E-04    (   1.8E-5  )   (   -1.6E-5 )   &   1.89E-04    (   3.5E-5  )   (   -3.3E-5 )   &   1.51E-04    (   2.1E-5  )   (   -2.1E-5 )   &   6.40E-05    (   1.7E-5  )   (   -1.7E-5 )   \\
icf(C)              &   1.03    (   0.04    )   (   -0.05   )   &&  1.20    (   0.01    )   (   -0.01   )   &   1.00    (   0.00    )   (   0.00    )   &   1.00    (   0.00    )   (   0.00    )   &   1.00    (   0.00    )   (   0.00    )   \\
C/H                 &   6.69E-04    (   2.7E-4  )   (   -2.7E-4 )   &&  3.53E-04    (   3.8E-5  )   (   -3.8E-5 )   &   1.89E-04    (   3.5E-5  )   (   -3.3E-5 )   &   4.03E-04    (   5.6E-5  )   (   -5.7E-5 )   &   6.40E-05    (   1.7E-5  )   (   -1.7E-5 )   \\
\\
N$^{+}$/H           &   1.49E-05    (   3.2E-6  )   (   -2.6E-6 )   &&  9.12E-07    (   2.7E-7  )   (   -2.1E-7 )   &   8.89E-06    (   1.8E-6  )   (   -1.5E-6 )   &   2.04E-06    (   2.4E-7  )   (   -2.2E-7 )   &   4.03E-06    (   7.8E-7  )   (   -5.9E-7 )   \\
icf(N)              &   10.37   (   2.69    )   (   -2.23   )   &&  58.03   (   22.43   )   (   -17.07  )   &   10.06   (   2.03    )   (   -1.70   )   &   48.73   (   11.41   )   (   -9.25   )   &   21.83   (   6.88    )   (   -6.70   )   \\
N$^{}$/H            &   1.55E-04    (   2.3E-5  )   (   -2.1E-5 )   &&  5.24E-05    (   1.5E-5  )   (   -1.2E-5 )   &   8.96E-05    (   1.2E-5  )   (   -1.2E-5 )   &   9.93E-05    (   2.1E-5  )   (   -1.8E-5 )   &   8.64E-05    (   2.1E-5  )   (   -2.0E-5 )   \\
\\
O$^{+}$/H           &   4.01E-05    (   1.4E-5  )   (   -9.4E-6 )   &&  4.25E-06    (   1.6E-6  )   (   -1.0E-6 )   &   3.30E-05    (   7.0E-6  )   (   -5.5E-6 )   &   5.76E-06    (   8.5E-7  )   (   -7.4E-7 )   &   1.41E-05    (   6.2E-6  )   (   -3.2E-6 )   \\
O$^{2+}$/H          &   2.76E-04    (   8.5E-6  )   (   -8.1E-6 )   &&  2.38E-04    (   4.3E-5  )   (   -3.6E-5 )   &   2.12E-04    (   1.6E-5  )   (   -1.5E-5 )   &   2.42E-04    (   4.2E-5  )   (   -3.6E-5 )   &   2.17E-04    (   2.3E-5  )   (   -2.1E-5 )   \\
icf(O)              &   1.65    (   0.04    )   (   -0.03   )   &&  1.14    (   0.01    )   (   -0.01   )   &   1.01    (   0.00    )   (   0.00    )   &   1.03    (   0.00    )   (   0.00    )   &   1.01    (   0.00    )   (   0.00    )   \\
O$^{}$/H            &   5.24E-04    (   2.8E-5  )   (   -2.3E-5 )   &&  2.77E-04    (   4.8E-5  )   (   -4.1E-5 )   &   2.47E-04    (   1.7E-5  )   (   -1.6E-5 )   &   2.55E-04    (   4.3E-5  )   (   -3.7E-5 )   &   2.35E-04    (   2.7E-5  )   (   -2.2E-5 )   \\
\\
Ne$^{2+}$/H         &   6.29E-05    (   3.9E-6  )   (   -3.7E-6 )   &&  5.72E-05    (   5.9E-6  )   (   -5.4E-6 )   &   7.88E-05    (   4.0E-6  )   (   -3.8E-6 )   &   6.45E-05    (   6.9E-6  )   (   -6.2E-6 )   &   5.82E-05    (   7.3E-6  )   (   -6.6E-6 )   \\
icf(Ne)             &   1.68    (   0.05    )   (   -0.04   )   &&  1.14    (   0.01    )   (   -0.01   )   &   1.27    (   0.04    )   (   -0.03   )   &   1.03    (   0.00    )   (   0.00    )   &   1.16    (   0.03    )   (   -0.03   )   \\
Ne$^{}$/H           &   1.06E-04    (   7.4E-6  )   (   -6.9E-6 )   &&  6.54E-05    (   6.7E-6  )   (   -6.1E-6 )   &   9.99E-05    (   5.0E-6  )   (   -4.7E-6 )   &   6.67E-05    (   7.1E-6  )   (   -6.4E-6 )   &   6.78E-05    (   9.3E-6  )   (   -7.7E-6 )   \\
\\
Ar$^{2+}$/H         &                               &&  2.63E-07    (   7.8E-8  )   (   -6.8E-8 )   &                               &   4.78E-07    (   1.6E-7  )   (   -1.5E-7 )   &   8.17E-07    (   2.1E-7  )   (   -1.8E-7 )   \\
Ar$^{3+}$/H         &   1.14E-06    (   3.4E-8  )   (   -3.0E-8 )   &&  7.03E-07    (   6.7E-8  )   (   -6.1E-8 )   &   3.04E-07    (   1.8E-8  )   (   -1.7E-8 )   &   5.14E-07    (   4.9E-8  )   (   -4.5E-8 )   &   4.43E-07    (   4.3E-8  )   (   -3.8E-8 )   \\
icf(Ar)             &   1.00    (   0.00    )   (   0.00    )   &&  1.00    (   0.00    )   (   0.00    )   &   1.00    (   0.00    )   (   0.00    )   &   1.00    (   0.00    )   (   0.00    )   &   1.00    (   0.08    )   (   0.00    )   \\
Ar$^{}$/H           &   1.14E-06    (   3.4E-8  )   (   -3.0E-8 )   &&  9.65E-07    (   1.3E-7  )   (   -1.1E-7 )   &   3.04E-07    (   1.8E-8  )   (   -1.7E-8 )   &   9.88E-07    (   2.0E-7  )   (   -1.7E-7 )   &   1.30E-06    (   2.9E-7  )   (   -2.4E-7 )   \\
\\
S$^{+}$/H           &   3.60E-07    (   7.8E-8  )   (   -5.5E-8 )   &&  9.06E-08    (   2.5E-8  )   (   -1.8E-8 )   &   7.17E-07    (   1.3E-7  )   (   -1.0E-7 )   &   1.50E-07    (   2.2E-8  )   (   -1.9E-8 )   &   3.12E-07    (   1.1E-7  )   (   -5.1E-8 )   \\
S$^{2+}$/H          &   2.61E-06    (   1.7E-7  )   (   -1.6E-7 )   &&  1.22E-06    (   2.2E-7  )   (   -1.8E-7 )   &   4.29E-06    (   4.7E-7  )   (   -4.2E-7 )   &   1.85E-06    (   2.6E-7  )   (   -2.3E-7 )   &   2.06E-06    (   2.7E-7  )   (   -2.4E-7 )   \\
icf(S)              &   2.30    (   0.21    )   (   -0.20   )   &&  2.73    (   0.29    )   (   -0.23   )   &   1.37    (   0.09    )   (   -0.08   )   &   2.29    (   0.12    )   (   -0.11   )   &   1.78    (   0.13    )   (   -0.18   )   \\
S$^{}$/H            &   6.86E-06    (   6.6E-7  )   (   -5.9E-7 )   &&  3.59E-06    (   8.2E-7  )   (   -6.7E-7 )   &   6.87E-06    (   8.5E-7  )   (   -7.5E-7 )   &   4.59E-06    (   7.9E-7  )   (   -6.7E-7 )   &   4.23E-06    (   6.8E-7  )   (   -5.6E-7 )   \\
\\
Cl$^{2+}$/H         &   6.81E-08    (   1.0E-8  )   (   -9.7E-9 )   &&  4.94E-08    (   5.8E-9  )   (   -5.2E-9 )   &   8.64E-08    (   2.2E-8  )   (   -2.2E-8 )   &   3.44E-08    (   4.3E-9  )   (   -3.8E-9 )   &   5.10E-08    (   9.6E-9  )   (   -9.6E-9 )   \\
icf(Cl)             &   2.61    (   0.15    )   (   -0.14   )   &&  1.00    (   0.00    )   (   0.00    )   &   1.56    (   0.06    )   (   -0.05   )   &   1.00    (   0.00    )   (   0.00    )   &   1.78    (   0.11    )   (   -0.78   )   \\
Cl$^{}$/H           &   1.78E-07    (   2.8E-8  )   (   -2.7E-8 )   &&  4.95E-08    (   5.9E-9  )   (   -5.2E-9 )   &   1.35E-07    (   3.6E-8  )   (   -3.5E-8 )   &   3.44E-08    (   4.3E-9  )   (   -3.9E-9 )   &   7.84E-08    (   2.9E-8  )   (   -2.1E-8 )   \\

\hline
\end{tabular}}
 \end{table*}

\begin{table*}
\centering \caption{Ionic and total abundances of IC 2553 and PB6.}
\label{TableA4} \scalebox{0.65}{
\begin{tabular}{lccccccc}

\hline
    &  \multicolumn{3}{c}{IC 2553}  && \multicolumn{2}{c}{PB6}\\    \cline{2-4} \cline{6-7}
\\
                        &       Entire PN                               &     K$_{\textrm{NE}}$                                        &       K$_{\textrm{SW}}$                                     &&       Entire PN                              &       K$_{\textrm{SE}}$ Knot       \\
\hline
\\
He$^{+}$/H          &   8.80E-02    (   6.0E-3  )   (   -5.6E-3 )   &   7.02E-02    (   4.3E-3  )   (   -4.3E-3 )   &   1.01E-01    (   6.5E-3  )   (   -6.1E-3 )   &&  4.67E-02    (   3.2E-3  )   (   -3.2E-3 )   &   6.64E-02    (   5.4E-3  )   (   -5.0E-3 )   \\
He$^{2+}$/H         &   2.09E-02    (   3.5E-4  )   (   -3.5E-4 )   &   1.25E-02    (   1.1E-3  )   (   -1.1E-3 )   &   8.09E-03    (   2.4E-4  )   (   -2.4E-4 )   &&  1.20E-01    (   1.1E-3  )   (   -1.1E-3 )   &   8.62E-02    (   2.5E-3  )   (   -2.5E-3 )   \\
He/H                &   1.09E-01    (   5.9E-3  )   (   -5.6E-3 )   &   8.27E-02    (   4.4E-3  )   (   -4.4E-3 )   &   1.09E-01    (   6.4E-3  )   (   -6.1E-3 )   &&  1.67E-01    (   3.2E-3  )   (   -3.2E-3 )   &   1.53E-01    (   5.7E-3  )   (   -5.5E-3 )   \\
\\
C$^{2+}$/H          &   6.17E-04    (   1.9E-5  )   (   -1.9E-5 )   &   6.04E-04    (   1.2E-4  )   (   -1.2E-4 )   &   5.48E-04    (   1.1E-4  )   (   -1.1E-4 )   &&  4.52E-04    (   9.6E-5  )   (   -9.2E-5 )   &                               \\
C$^{3+}$/H          &   6.20E-05    (   1.2E-5  )   (   -1.3E-5 )   &                               &                               &&  2.74E-04    (   4.4E-5  )   (   -4.4E-5 )   &                               \\
icf(C)              &   1.19    (   0.01    )   (   -0.02   )   &   1.09    (   0.02    )   (   -0.03   )   &   1.07    (   0.03    )   (   -0.03   )   &&  1.00    (   0.00    )   (   0.00    )   &                               \\
C/H                 &   8.06E-04    (   3.0E-5  )   (   -3.0E-5 )   &   6.59E-04    (   1.6E-4  )   (   -1.6E-4 )   &   5.88E-04    (   1.4E-4  )   (   -1.4E-4 )   &&  7.26E-04    (   1.1E-4  )   (   -1.0E-4 )   &                               \\
\\
N$^{+}$/H           &   3.84E-06    (   9.8E-7  )   (   -6.3E-7 )   &   1.22E-05    (   1.5E-6  )   (   -1.4E-6 )   &   1.21E-05    (   1.5E-6  )   (   -1.3E-6 )   &&  2.51E-05    (   2.3E-6  )   (   -2.1E-6 )   &   4.12E-05    (   5.3E-6  )   (   -4.7E-6 )   \\
icf(N)              &   45.06   (   14.66   )   (   -14.93  )   &   12.69   (   2.87    )   (   -2.23   )   &   11.77   (   2.68    )   (   -2.18   )   &&  15.04   (   1.86    )   (   -1.66   )   &   8.99    (   1.79    )   (   -1.49   )   \\
N$^{}$/H            &   1.70E-04    (   4.6E-5  )   (   -5.0E-5 )   &   1.55E-04    (   2.9E-5  )   (   -2.4E-5 )   &   1.42E-04    (   2.6E-5  )   (   -2.2E-5 )   &&  3.77E-04    (   2.9E-5  )   (   -2.7E-5 )   &   3.70E-04    (   5.3E-5  )   (   -4.6E-5 )   \\
\\
O$^{+}$/H           &   7.70E-06    (   3.7E-6  )   (   -1.8E-6 )   &   2.76E-05    (   5.4E-6  )   (   -4.7E-6 )   &   3.07E-05    (   6.3E-6  )   (   -5.1E-6 )   &&  1.74E-05    (   2.4E-6  )   (   -2.1E-6 )   &   2.82E-05    (   6.0E-6  )   (   -4.9E-6 )   \\
O$^{2+}$/H          &   3.33E-04    (   2.4E-5  )   (   -2.2E-5 )   &   3.11E-04    (   3.8E-5  )   (   -3.4E-5 )   &   3.01E-04    (   3.5E-5  )   (   -3.1E-5 )   &&  1.14E-04    (   3.3E-6  )   (   -3.2E-6 )   &   1.50E-04    (   1.8E-5  )   (   -1.6E-5 )   \\
icf(O)              &   1.13    (   0.01    )   (   -0.01   )   &   1.10    (   0.01    )   (   -0.01   )   &   1.04    (   0.00    )   (   0.00    )   &&  2.62    (   0.12    )   (   -0.12   )   &   1.78    (   0.07    )   (   -0.07   )   \\
O$^{}$/H            &   3.91E-04    (   3.1E-5  )   (   -2.8E-5 )   &   3.73E-04    (   4.3E-5  )   (   -3.9E-5 )   &   3.47E-04    (   3.7E-5  )   (   -3.3E-5 )   &&  3.44E-04    (   2.0E-5  )   (   -1.7E-5 )   &   3.19E-04    (   3.3E-5  )   (   -3.0E-5 )   \\
\\
Ne$^{2+}$/H         &   9.08E-05    (   4.4E-6  )   (   -4.4E-6 )   &   1.17E-04    (   1.9E-5  )   (   -1.7E-5 )   &   1.19E-04    (   1.6E-5  )   (   -1.4E-5 )   &&  3.21E-05    (   9.4E-7  )   (   -9.2E-7 )   &   4.64E-05    (   6.5E-6  )   (   -5.7E-6 )   \\
icf(Ne)             &   1.13    (   0.01    )   (   -0.01   )   &   1.10    (   0.01    )   (   -0.01   )   &   1.05    (   0.00    )   (   0.00    )   &&  3.12    (   0.23    )   (   -0.22   )   &   1.85    (   0.10    )   (   -0.09   )   \\
Ne$^{}$/H           &   1.03E-04    (   5.3E-6  )   (   -5.0E-6 )   &   1.29E-04    (   2.1E-5  )   (   -1.8E-5 )   &   1.24E-04    (   1.7E-5  )   (   -1.5E-5 )   &&  9.96E-05    (   8.7E-6  )   (   -7.1E-6 )   &   8.56E-05    (   1.2E-5  )   (   -1.0E-5 )   \\
\\
Ar$^{2+}$/H         &   1.07E-06    (   9.8E-8  )   (   -9.0E-8 )   &   1.20E-06    (   2.3E-7  )   (   -1.9E-7 )   &                               &&                              &                               \\
Ar$^{3+}$/H         &   1.36E-06    (   6.0E-8  )   (   -5.7E-8 )   &   6.76E-07    (   8.6E-8  )   (   -7.6E-8 )   &   5.78E-07    (   7.7E-8  )   (   -6.8E-8 )   &&  8.99E-07    (   3.2E-8  )   (   -2.7E-8 )   &   9.06E-07    (   1.2E-7  )   (   -1.0E-7 )   \\
icf(Ar)             &   1.00    (   0.00    )   (   0.00    )   &   1.04    (   0.06    )   (   -0.04   )   &   1.00    (   0.00    )   (   0.00    )   &&  1.00    (   0.00    )   (   0.00    )   &   1.00    (   0.00    )   (   0.00    )   \\
Ar$^{}$/H           &   2.42E-06    (   1.5E-7  )   (   -1.4E-7 )   &   1.98E-06    (   3.9E-7  )   (   -3.2E-7 )   &   5.78E-07    (   7.7E-8  )   (   -6.8E-8 )   &&  8.99E-07    (   3.2E-8  )   (   -2.7E-8 )   &   9.06E-07    (   1.2E-7  )   (   -1.0E-7 )   \\
\\
S$^{+}$/H           &   2.61E-07    (   1.1E-7  )   (   -3.7E-8 )   &   5.54E-07    (   6.7E-8  )   (   -5.7E-8 )   &   7.26E-07    (   9.6E-8  )   (   -8.0E-8 )   &&  3.61E-07    (   3.4E-8  )   (   -3.1E-8 )   &   5.23E-07    (   6.1E-8  )   (   -5.4E-8 )   \\
S$^{2+}$/H          &   3.71E-06    (   4.9E-7  )   (   -4.3E-7 )   &   2.87E-06    (   5.0E-7  )   (   -4.2E-7 )   &   4.39E-06    (   7.6E-7  )   (   -6.5E-7 )   &&  1.77E-06    (   9.2E-8  )   (   -9.2E-8 )   &   1.70E-06    (   3.1E-7  )   (   -2.6E-7 )   \\
icf(S)              &   2.53    (   0.18    )   (   -0.22   )   &   1.78    (   0.11    )   (   -0.11   )   &   1.62    (   0.11    )   (   -0.10   )   &&  3.59    (   0.23    )   (   -0.21   )   &   2.30    (   0.18    )   (   -0.16   )   \\
S$^{}$/H            &   9.96E-06    (   1.7E-6  )   (   -1.5E-6 )   &   6.10E-06    (   1.1E-6  )   (   -9.4E-7 )   &   8.30E-06    (   1.6E-6  )   (   -1.3E-6 )   &&  7.64E-06    (   5.5E-7  )   (   -5.2E-7 )   &   5.13E-06    (   8.8E-7  )   (   -7.5E-7 )   \\
\\
Cl$^{2+}$/H         &   8.33E-08    (   8.0E-9  )   (   -7.3E-9 )   &   7.61E-08    (   9.4E-9  )   (   -8.3E-9 )   &   7.94E-08    (   9.6E-9  )   (   -8.6E-9 )   &&  4.03E-08    (   7.9E-9  )   (   -7.4E-9 )   &   3.47E-08    (   5.1E-9  )   (   -4.4E-9 )   \\
icf(Cl)             &   1.00    (   0.00    )   (   0.00    )   &   1.90    (   0.09    )   (   -0.08   )   &   1.76    (   0.08    )   (   -0.07   )   &&  4.09    (   0.21    )   (   -0.20   )   &   2.69    (   0.14    )   (   -0.13   )   \\
Cl$^{}$/H           &   8.46E-08    (   1.2E-8  )   (   -8.0E-9 )   &   1.45E-07    (   2.1E-8  )   (   -2.0E-8 )   &   1.40E-07    (   2.1E-8  )   (   -1.8E-8 )   &&  1.65E-07    (   3.6E-8  )   (   -3.3E-8 )   &   9.35E-08    (   1.6E-8  )   (   -1.4E-8 )   \\

\hline
\end{tabular}}
 \end{table*}

\end{appendix}


\end{document}